\PassOptionsToPackage{table,xcdraw}{xcolor}

\documentclass[sigconf]{acmart}

\renewcommand\footnotetextcopyrightpermission[1]{} 
\newcommand\new[1]{#1} 

\usepackage[english]{babel}
\usepackage{blindtext}
\usepackage{balance}

\usepackage[absolute,showboxes]{textpos}

\usepackage{tikz}
\newcommand*\circled[1]{\tikz[baseline=(char.base)]{
            \node[shape=circle,draw,inner sep=2pt] (char) {#1};}}

\usepackage{paralist}

\usepackage{soul}
\vspace{-0.71em}

\usepackage{placeins}

\newcommand{\eat}[1]{}
\newcommand{\numdevices}{96\xspace}
\newcommand{\numproducts}{56\xspace}
\newcommand{\numvendors}{40\xspace}
\newcommand{\numexperiments}{9,810\xspace}

\usepackage{tabularx}
\newcolumntype{L}[1]{>{\raggedright\arraybackslash}p{#1}}
\newcolumntype{C}[1]{>{\centering\arraybackslash}p{#1}}
\newcolumntype{R}[1]{>{\raggedleft\arraybackslash}p{#1}}

\usepackage{algorithm}
\usepackage[noend]{algorithmic}

\usepackage{enumitem}
\setlist{nolistsep}
\usepackage{graphicx}
\usepackage{epstopdf}
\usepackage{grffile}
\usepackage{amsmath}
\usepackage{multicol,multirow}
\usepackage{rotating,url}
\usepackage{enumitem}
\usepackage{color}
\usepackage{subfigure}
\usepackage{xspace}
\usepackage{ifpdf}
\usepackage{mdwlist}
\usepackage{colortbl}
\usepackage{caption}
\usepackage{booktabs}
\usepackage{multirow}
\usepackage{enumitem}
\usepackage{xfrac}
\usepackage{afterpage}
\usepackage{pdflscape}
\usepackage{longtable}
\usepackage{hhline}
\usepackage{lscape}
\usepackage[export]{adjustbox}

\newlength{\oldtextfloatsep}
\newcommand{\ie}{i.e., \@}
\newcommand{\eg}{e.g., \@}

\usepackage{enumitem}
\usepackage{mdwlist}
\setlist{nolistsep}
\setlist[description]{noitemsep,topsep=0pt,parsep=0pt,partopsep=0pt,leftmargin=0pt}
\usepackage{pifont}

\usepackage{everypage}

\newcommand{\todo}[1]{\textcolor{red}{TODO: \emph{#1}}}
\newcommand{\tocut}[1]{\textcolor{blue}{\emph{#1}}}
\newcommand{\oldnotes}[1]{\textcolor{gray}{#1}}
\newcommand{\amm}[1]{\textcolor{magenta}{AMM: \emph{#1}}}
\newcommand{\djd}[1]{\textcolor{teal}{DJD: \emph{#1}}}
\newcommand{\rk}[1]{\textcolor{purple}{RK: \emph{#1}}}
\newcommand{\drc}[1]{\textcolor{olive}{[[DC: \emph{#1}]]}}
\eat{
\newcommand{\todo}[1]{}
\newcommand{\tocut}[1]{}
\newcommand{\oldnotes}[1]{}
\newcommand{\amm}[1]{}
\newcommand{\djd}[1]{}
\newcommand{\rk}[1]{}
\newcommand{\drc}[1]{}
}

\newcommand{\para}[1]{\vspace{0.1em} \noindent \textbf{#1}}
\long\def\comment#1{}

\begin{document}


\clubpenalty=10000 
\widowpenalty = 10000

\title[A Haystack Full of Needles:\\Scalable Detection of IoT Devices in the Wild]{A Haystack Full of Needles:\\Scalable Detection of IoT Devices in the Wild}
%
\author{Said Jawad Saidi}
\affiliation{%
	\institution{Max Planck Institute for Informatics}
}

\author{Anna Maria Mandalari}
\affiliation{%
	\institution{Imperial College London }
}

\author{Roman Kolcun}
\affiliation{%
	\institution{Imperial College London }
}

\author{Hamed Haddadi}
\affiliation{%
	\institution{Imperial College London }}

\author{Daniel J. Dubois}
\affiliation{%
	\institution{Northeastern University}
}

\author{David Choffnes}
\affiliation{%
	\institution{Northeastern University }}

\author{Georgios Smaragdakis}
\affiliation{%
		\institution{TU Berlin\\Max Planck Institute for Informatics}}

\author{Anja Feldmann}
\affiliation{\institution{Max Planck Institute for Informatics/Saarland University }}

\renewcommand{\shortauthors}{Saidi et al.} 

\begin{abstract}


Consumer Internet of Things (IoT) devices are extremely popular, providing users with rich and diverse functionalities, from voice assistants to home appliances. 
These functionalities often come with significant privacy and security risks, with notable recent large-scale coordinated global attacks disrupting large service providers. 
Thus, an important first step to address these risks is to know \emph{what} IoT devices are \emph{where} in a network. 
While some limited solutions exist, a key question is whether device discovery can be done by Internet service providers that only see sampled flow statistics. 
In particular, it is challenging for an ISP to efficiently and effectively track and trace activity from IoT devices deployed by its millions of subscribers---all with sampled network data.

In this paper, we develop and evaluate a scalable methodology to accurately detect and monitor IoT devices at subscriber lines with limited, highly sampled data in-the-wild. 
Our findings indicate that millions of IoT devices are detectable and identifiable within hours, both at a major ISP as well as an IXP, using \emph{passive}, sparsely \emph{sampled} network flow headers.
Our methodology is able to detect devices from more than 77\% of the studied IoT manufacturers, including popular devices such as smart speakers.  
While our methodology is effective for providing network analytics, it also highlights significant privacy consequences.

\end{abstract}

\ccsdesc[300]{Security and privacy~Network security}
\ccsdesc[300]{Networks~Network monitoring}
\ccsdesc[500]{Networks~Public Internet}
\ccsdesc[500]{Networks~Network measurement}

\keywords{Internet of Things, IoT detection, IoT secuirty and privacy, Internet Measurement}

\setlength{\TPHorizModule}{\paperwidth}
\setlength{\TPVertModule}{\paperheight}
\TPMargin{5pt}
\begin{textblock}{0.8}(0.1,0.02)
  \noindent
  \footnotesize
  If you cite this paper, please use the IMC reference:
  Said Jawad Saidi, Anna Maria Mandalari, Roman Kolcun, Hamed Haddadi, Daniel J. Dubois, David Choffnes, Georgios Smaragdakis, Anja Feldmann. 2020.
  A Haystack Full of Needles: Scalable Detection of IoT Devices in the Wild.
  In \textit{Internet Measurement Conference (IMC '20), October 27--29, 2020, Virtual Event, USA.}
  ACM, New York, NY, USA, 14 pages. https://doi.org/10.1145/3419394.3423650
\end{textblock}

\maketitle

\section{Introduction}\label{sec:intro}

The number of IoT devices deployed within homes is increasing rapidly. 
It is estimated that at the end of 2019, more than 9.5 billion IoT devices were active, and the IoT population will increase to 20 billion by 2025~\cite{IoT-stats}. 
Such devices include virtual assistants, smart home control, cameras, and smart TVs.
While users deploy some IoT devices explicitly, they are often unaware of the security threats and privacy consequences of using such devices~\cite{Deep-Insecurities-CACM2019}. 
Major Internet Service Providers (ISPs) are developing strategies for dealing with the large-scale coordinated attacks from these devices.

Existing solutions focus on instrumenting testbeds or home environments to collect and analyze full packet captures~\cite{PETS20_DuboisKMPCH20, Information-Exposure-IMC2019, IMC19_DekovenRMABSSVS19}, local search for IoT anomalies~\cite{AUDI-IoT-JSAC,IoT-All-Things-2019}, active measurements~\cite{ZMap,CCS15_DurumericAMBH15}, or data from antivirus companies running scan campaigns from users homes~\cite{IoT-All-Things-2019}. 
In isolation, these data sources do not provide enough insights for preventing network-wide attacks from IoT devices~\cite{haddadi2018siotome}.
Detecting IoT devices from an ISP can help to identify suspicious traffic and what devices are common among the subscriber lines generating that traffic.

In this paper, we present a methodology for detecting home IoT devices in-the-wild at an ISP, and an Internet Exchange Point (IXP), by relying on passive, sampled network traces and active probing experiments. 
We build on the insight that IoT devices typically rely on backend infrastructure hosted on the cloud to offer their services. 
While contacting such infrastructure, they expose information, including their traffic destinations, even when a device is not in use~\cite{Information-Exposure-IMC2019}.  
One of the challenges of detecting IoT devices at scale is the \emph{poor availability and low granularity} of data sources.
The available data is often in the form of centrally-collected aggregate and sampled data (\eg NetFlow~\cite{rfc3954}, IPFIX traces~\cite{rfc7011}). 
Thus, we need a methodology that (a) does not rely on payload and (b) handles sparsely sampled data.

Another challenge is \emph{traffic patterns diversity}, across IoT devices and their services.\footnote{Here we refer to IoT services as the set of protocols and destinations that are part of the operations of an IoT device.} 
We note that some devices, e.g., cameras, will generate significant continuous traffic; others, e.g., plugs, can be expected to be mainly passive unless used. 
Moreover, many devices offer the same service, e.g., the Alexa voice assistant~\cite{alexavoiceservice} is available on several brands of smart speakers as well as on Amazon Fire TV devices.
Here, the traffic patterns may depend on the service rather than the specific IoT device.
Some services rely on dedicated backend infrastructures, while others may use shared ones, \eg~CDNs. 
Thus, we need a methodology that identifies which IoT services are detectable from the traffic and then identifies a unique traffic pattern for each IoT device and associated services.

Our key insight is that we can address these challenges by focusing our analysis only on the types of destinations contacted by IoT devices. 
Even with sparsely sampled data, the set of servers contacted by an IoT device over time can form a reasonably unique signature that is revealed in as little as a few hours. 
However, this approach has limitations, for example we cannot use it to detect devices or services that use a shared infrastructure with unrelated services (\eg CDNs).

\begin{figure}[t]
\captionsetup{skip=0.35em,font=small}
  \centering
  \includegraphics[width=0.8\linewidth]{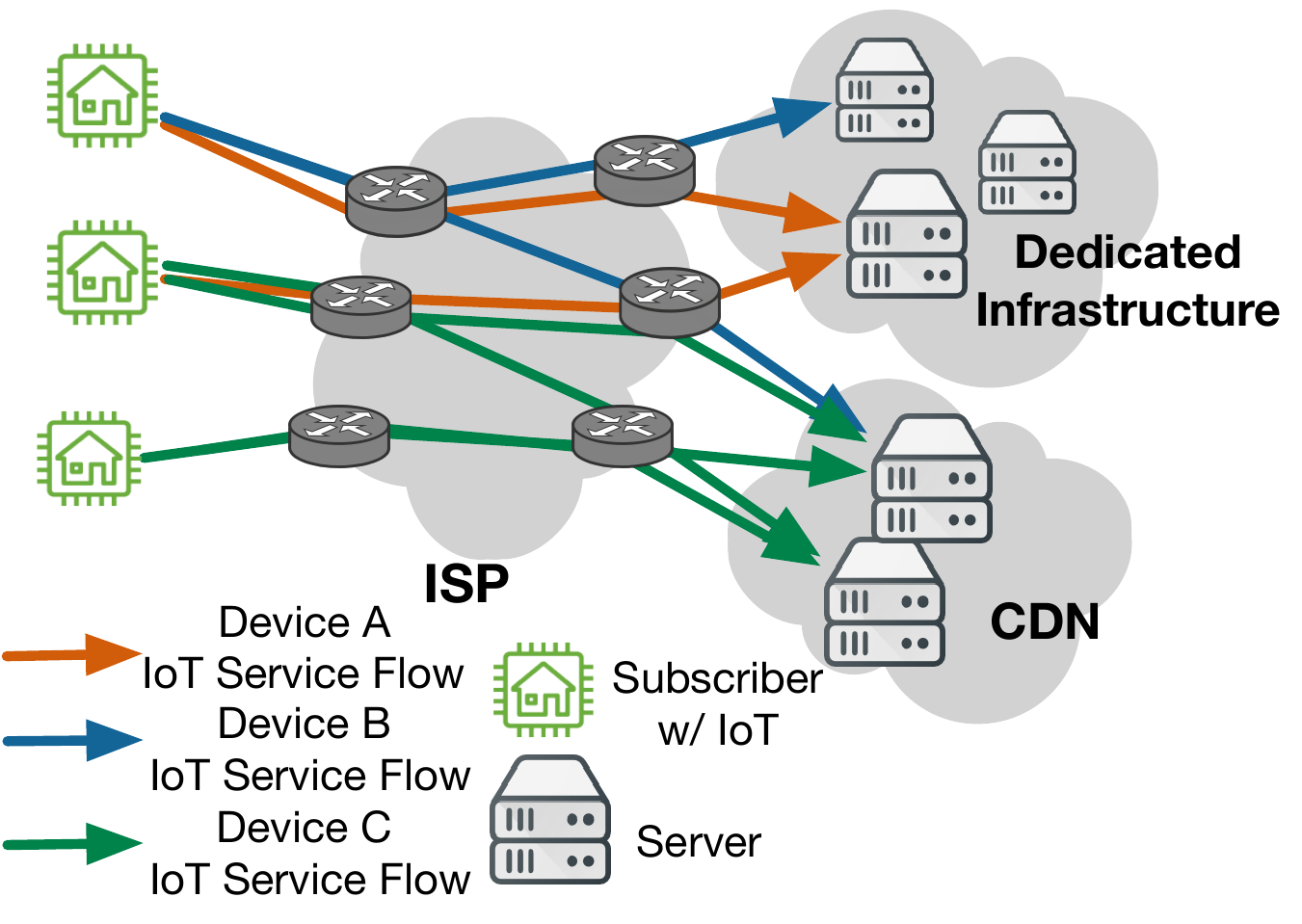}
\caption{Simplified IoT communication patterns.} 
\label{fig:iot-communication-pattern}
\end{figure}

To understand the detectability of IoT devices in the above-mentioned environment, we focus on the possible communication patterns of end-user IoT services and the types of destinations they contact.
Figure~\ref{fig:iot-communication-pattern} shows three possible communication patterns on top of a typical network topology. 
This includes three households, an ISP, as well as a dedicated infrastructure and a CDN that hosts multiple servers. 
Device \emph{A} is deployed by two subscribers, and only contacts one server in the dedicated infrastructure. 
Device \emph{B} is deployed by a single subscriber and contacts both a dedicated server, as well as a CDN server. 
Device \emph{C} is deployed by two subscribers and contacts only CDN servers. 
We observe that, using NetFlow traces at the ISP edge, it is possible to identify subscriber lines hosting devices of type \emph{A} and \emph{B}. 
Devices of type \emph{C} are harder to detect given the sampling rates and header-only nature of NetFlow. 

\begin{figure*}[t]

 \centering
  \begin{minipage}[t]{0.33\linewidth}
    \subfigure{
    \includegraphics[width=0.9\linewidth]{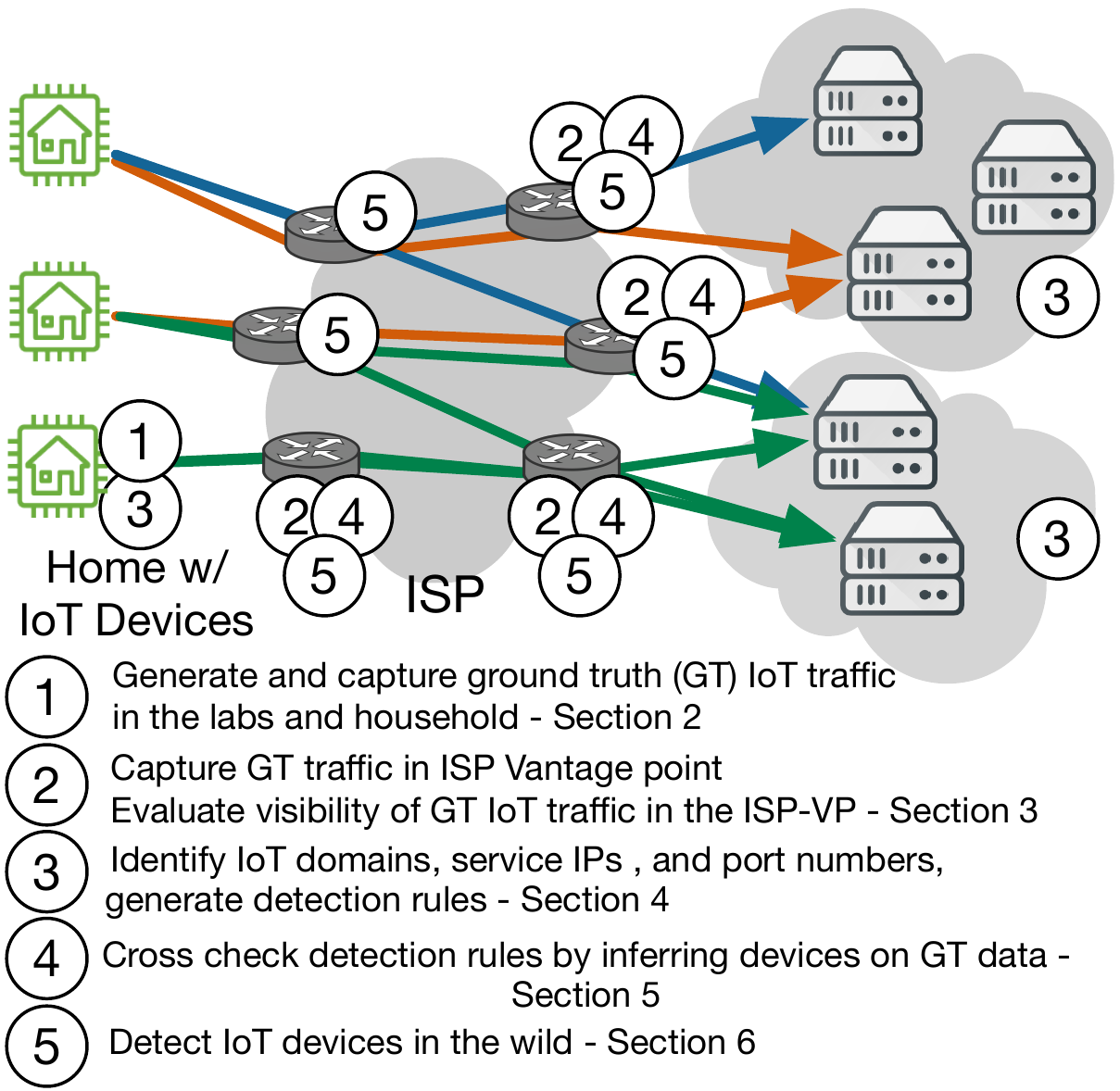}
    }
    \vspace{-0.5cm}
      \centering
     \captionsetup{justification=centering,font=small}
   \caption{General methodology overview.}
      \label{fig:general-overview}
  \end{minipage}
  \hfill
  \begin{minipage}[t]{0.33\linewidth}
    \centering
\subfigure{
  \includegraphics[width=0.9\linewidth]{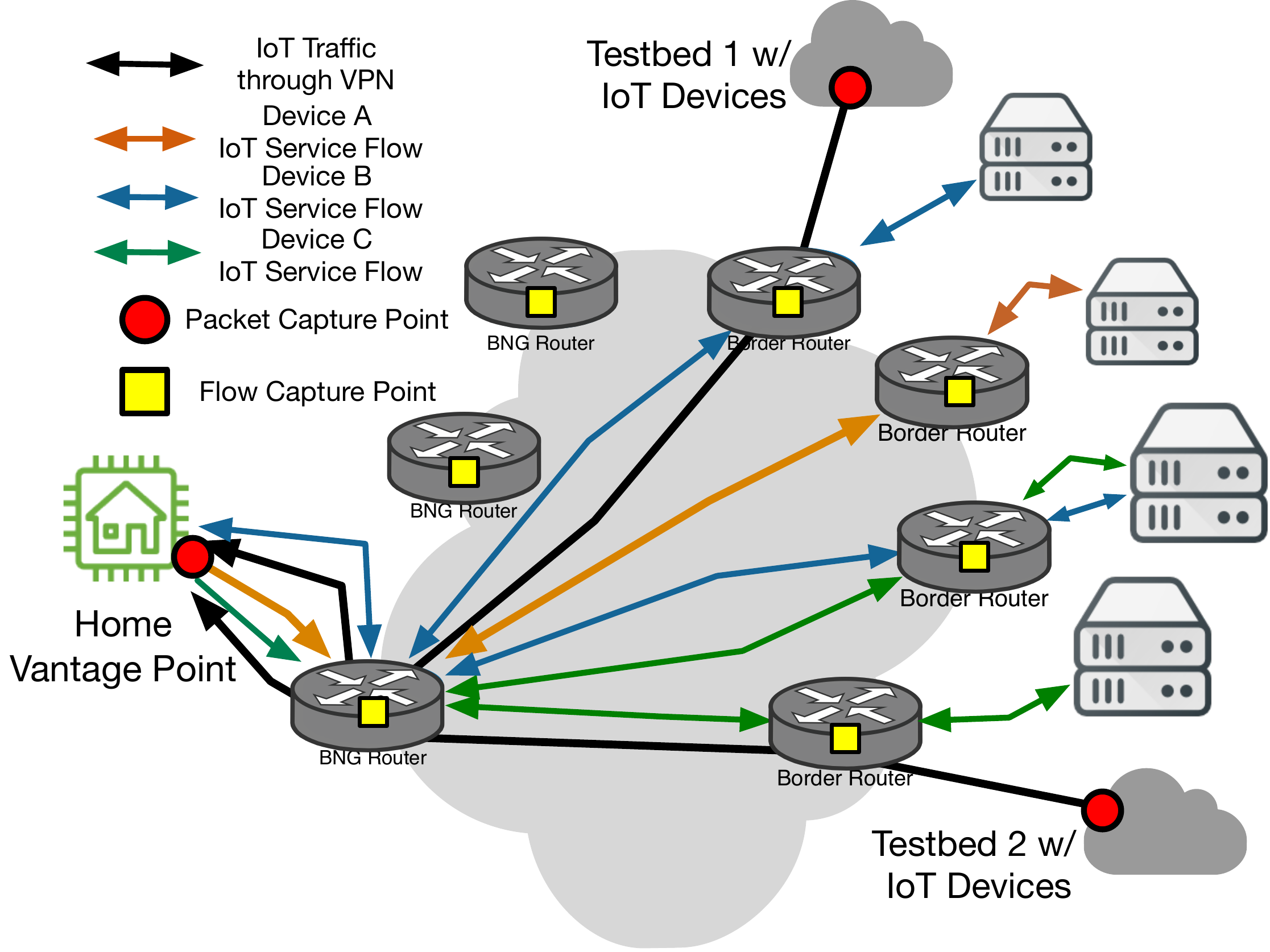}
  }
  \centering
\captionsetup{justification=centering,font=small}
    \caption{ISP setup \\ \& flow collection points.} 
        \label{fig:ISP-measurement}
  \end{minipage}
 \hfill
  \begin{minipage}[t]{0.33\linewidth}
    \subfigure{
      \includegraphics[width=1\linewidth]{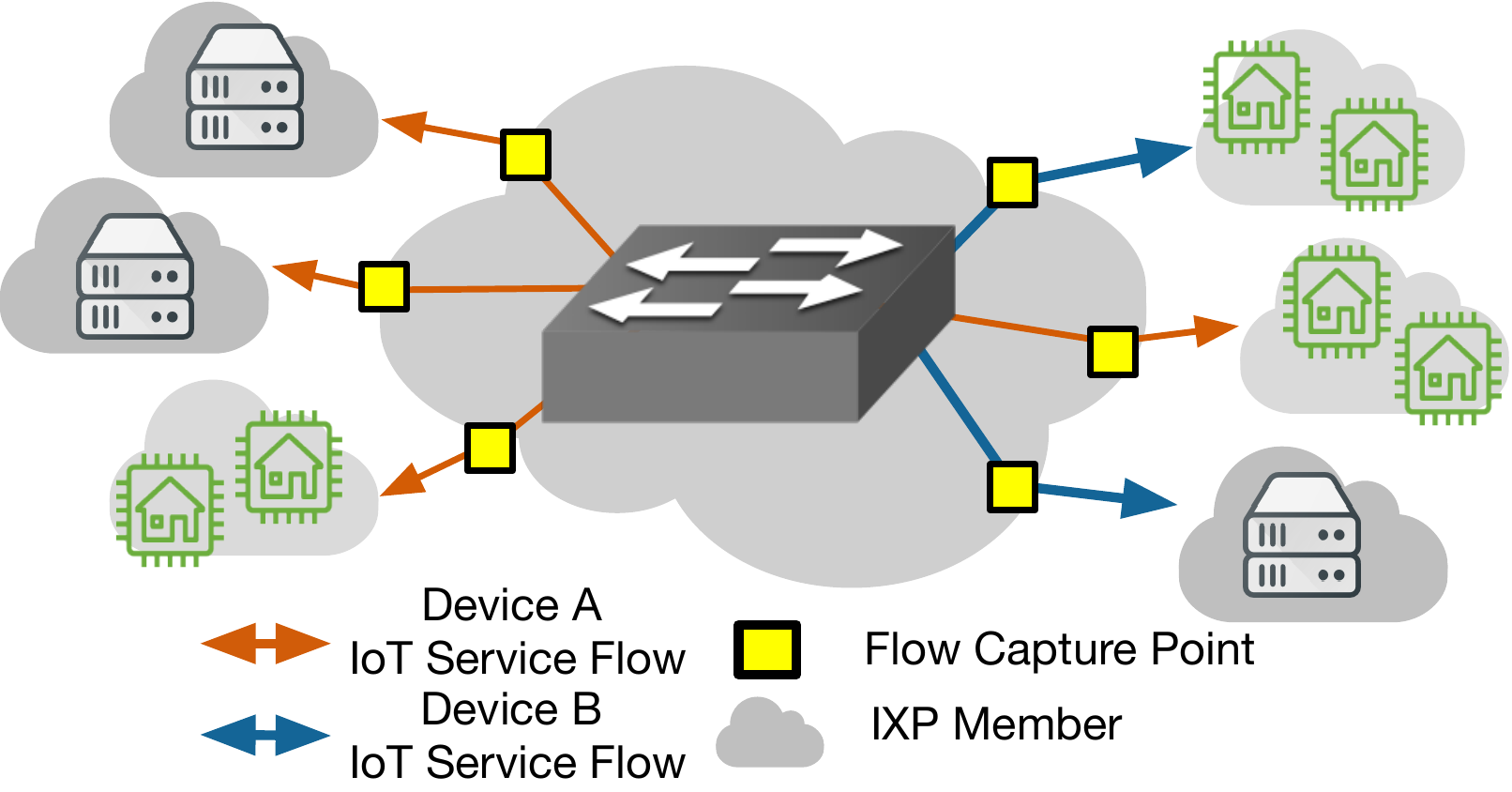}
    }
    \vspace{0.5cm}
      \centering
\captionsetup{justification=centering,font=small}
    \caption{IXP setup \\ \& flow collection points.} 
        \label{fig:IXP-measurement}
  \end{minipage}
\vspace{1cm}
\end{figure*}

In this paper, we use a unique testbed and dataset to build a methodology for detecting and monitoring IoT devices at scale~(see Figure~\ref{fig:general-overview}). 
We first use controlled experiments, where we tunnel the traffic of two IoT testbeds with \numdevices{} IoT devices to an ISP. 
This provides us with ground truth IoT traffic within this ISP (Section~\ref{sec:IoT-traffic-subscriber}). 
We confirm the visibility of the ground truth IoT traffic using the NetFlow ISP data (Section~\ref{sec:visibility}). 
Next, we identify backend infrastructures for many IoT services, from the observed ISP IoT traffic (Section~\ref{sec:methodology}).  
We augment this base information with data from DNS queries, web certificates, and banners. 
Next, we use the traffic signatures to identify broadband subscriber lines using IoT services at the ISP, as well as an IXP (Section~\ref{sec:iot-wild-results}). 
Finally, we discuss 
our results, their significance, and limitations in Section~\ref{sec:discussion}, 
related work (Section~\ref{sec:related-work}), 
and conclude with a summary in Section~\ref{sec:conclusion}.

Our main contributions are as follows:  

\begin{itemize}[leftmargin=*]

\item We develop a methodology for identifying IoT devices, by classifying domains and IP addresses of the backend infrastructure.     
To this end we derive distinct signatures, in terms of IP/domain/port destinations, to recognize IoT devices. With our signatures we were able to recognize the presence of devices from 31 out of \numvendors manufacturers in our testbed.\footnote{To foster further research in the area of IoT privacy and security, we make all the signatures available at \url{https://moniotrlab.ccis.neu.edu/imc20/}}

\item We show that it is possible to detect the presence of IoT devices at subscriber lines, using sparsely sampled flow captures from a large residential ISP, and a major IXP, even if the device is idle, i.e., not in active use.
Specifically, we were able to recognize that 20\% of 15 million subscriber lines used at least one of the \numproducts different IoT products in our testbed.

\item We highlight that our technique scales, is accurate, and can identify millions of IoT devices within minutes, in a non-intrusive way from passive, sampled data. 
In the case of the ISP, we were able to detect the presence of devices from 72\% of our target manufacturers within 1 hour, sometimes minutes. 
\end{itemize}
Based on our findings, we also discuss why some IoT devices are faster to detect, how to hide an IoT service, as well as how the detectability can be used to improve IoT services and network troubleshooting.

\section{IoT -- Controlled Experiments}
\label{sec:IoT-traffic-subscriber} 

We need ground truth traffic from IoT devices, as observed both in a testbed and in the wild, for developing and testing our methodology. 
In this section, we describe our data collection strategy (see point \circled{1} of Figure~\ref{fig:general-overview}).

\subsection{Network Setting}

We utilize two \emph{vantage points}, namely a large European ISP, and a major European IXP.

\noindent\textbf{ISP (ISP-VP).} The ISP is a large residential ISP that offers Internet services to over 15 million broadband subscriber lines. 
The ISP uses NetFlow~\cite{rfc3954} to monitor the traffic flows at all border routers in its network, using a consistent sampling rate across all routers. Figure~\ref{fig:ISP-measurement} shows where NetFlow data is collected.

\noindent\textbf{IXP (IXP-VP).} 
The IXP facilitates traffic exchange between its members. At this point, it has more than $800$ members, including international, with peak traffic exceeding 8~Tbps. 
The IXP uses IPFIX~\cite{rfc7011} to collect traffic data across its switching
fabric at a consistent sampling rate, which is an order of magnitude lower than
the one used at the ISP.
Figure~\ref{fig:IXP-measurement} illustrates where the IPFIX data is collected.

\noindent\textbf{Ethical considerations ISP/IXP.} Neither the ISP nor the IXP flow data contain any payload data, thus no user information. 
We distinguish user IPs from server IPs and anonymize by hashing all user IPs, \new{following the method described
in~\cite{IMC19_DekovenRMABSSVS19}. The address space of the ISP residential users is known}. We call an IP a server IP if it receives or transmits traffic on well-known ports or
if it belongs to ASes of cloud or CDN providers. 
The ports include, e.g., web ports (80, 443, 8080), NTP (123), DNS (53). 
Moreover, we do not have any specific user activity and can only access and report aggregated statistics in accordance
with the policies of the ISP and IXP.

\noindent\textbf{Subscriber line (Home-VP) Network setup.}
 In order to ingest ground truth traffic into the network, we need privileged access to a \emph{home subscriber line}. 
For this, we use the ISP-VP, but rather than deploying all IoT devices directly
within the home, we placed a VPN endpoint with an IP out of the /28 subscriber's prefix and used it to ingest 
IoT traffic tunneled to the server from two IoT testbeds, one in Europe, one in the US, see Figure~\ref{fig:ISP-measurement}. 
The measurement points within the ISP will also capture this traffic. We simply excluded this traffic from our dataset, as the VPN tunnel endpoints are known to us and for each experiment we use the default DNS server for the ISP.  
Importantly, since the /28 prefix is used explicitly for our experiments, there was no other network activity other than that of the IoT devices.

\noindent\textbf{Ethical considerations--Home-VP setting.} 
With the cooperation of the ISP, we were able to use a reserved /28 allocated to this specific subscriber line (Home-VP) (with signed explicit consent) out of a /22 prefix reserved for residential users. 
Thus, the analysis in this paper only considers traffic explicitly ingested by the ground truth experiments and does not involve any user-generated traffic.

\subsection{Ground Truth Traffic Setting}

The IoT testbeds used here consist of \numdevices devices from \numvendors vendors.  
We selected the devices to provide diversity within and between different categories: surveillance, smart hubs, home automation, video, audio, and appliances. 
\new{Most of these are among the most popular devices, according to Amazon, in their respective region.} 
Our testbed includes multiple instances of the same device (\numproducts different products), so that we can see the destinations that each product contacts in different locations.
For a list of the IoT devices and the category of each device, we refer to Table~\ref{table:devices}.
We redirect all IoT traffic to the Home-VP within the ISP, and  
we  capture all the traffic generated by the IoT devices (see \circled{1} in Figure~\ref{fig:general-overview}). 
\begin{table}[!bpt]
  \captionsetup{skip=0.2em, font=small}
	\centering
	\rowcolors{3}{gray!10}{white}
		\resizebox{1.0\columnwidth}{!}{	
		\Huge
\begin{tabularx}{\textwidth}{l|X}
			\textbf{Category}& \textbf{Device Name}
			\\ \hline
\emph{Surveillance} & 
Amcrest~Cam, Blink~Cam, Blink~Hub, Icsee~Doorbell, Lefun~Cam, Luohe~Cam, 
Microseven~Cam, Reolink~Cam, Ring~Doorbell, Ubell~Doorbell, Wansview~Cam, Yi~Cam, ZModo Doorbell 
\\ \hline
\emph{Smart Hubs} &  
Insteon, Lightify, Philips~Hue, Sengled, Smartthings, SwitchBot, Wink~2, Xiaomi 
\\ \hline
\emph{Home Automation} &
D-Link~Mov~Sensor, Flux~Bulb, Honeywell~T-stat, Magichome~Strip, Meross~Door Opener, Nest~T-stat, Philips~Bulb, 
Smartlife~Bulb, Smartlife~Remote, TP-Link~Bulb, TP-Link~Plug, WeMo~Plug, Xiaomi~Strip, Xiaomi~Plug
\\ \hline
\emph{Video} &
Apple~TV, Fire~TV, LG~TV, Roku~TV, Samsung~TV 	
\\ \hline
\emph{Audio} & 
Allure~with~Alexa, Echo~Dot, Echo~Spot, Echo~Plus, Google~Home~Mini, Google~Home
\\ \hline
\emph{Appliances} & 
Anova~Sousvide,  Appkettle, GE Microwave, Netatmo~Weather, Samsung~Dryer (idle), Samsung~Fridge (idle), 
Smarter Brewer, Smarter~Coffee~Machine, Smarter~iKettle, Xiaomi~Rice~Cooker
\\ \hline
\end{tabularx}
}
\caption{\textbf{IoT devices under test.} \emph{idle} indicates that we capture the traffic just for idle periods because the experiments could not be automated.} 
\label{table:devices}
\end{table}

Most of the selected IoT devices are controlled using either a voice interface provided by a voice assistant (such as Amazon Alexa) or via a smartphone companion application. 
We use the voice interface to automate active experiments by producing voice commands using a Google Voice synthesizer.
For IoT devices that support a companion app, we use Android smartphones, and we rely on the Monkey Application Exerciser for Android Studio~\cite{monkey} for automating simulated interactions between the user and the IoT device. 

\begin{figure*} [t!]
 \captionsetup{skip=1em,width=0.5\linewidth,font=small}
 \centering
\begin{minipage}[t]{1\linewidth}
\vspace{-0.2cm}
  \subfigure[\# Unique service IPs per hour.]{
     \includegraphics[width=0.48\linewidth]{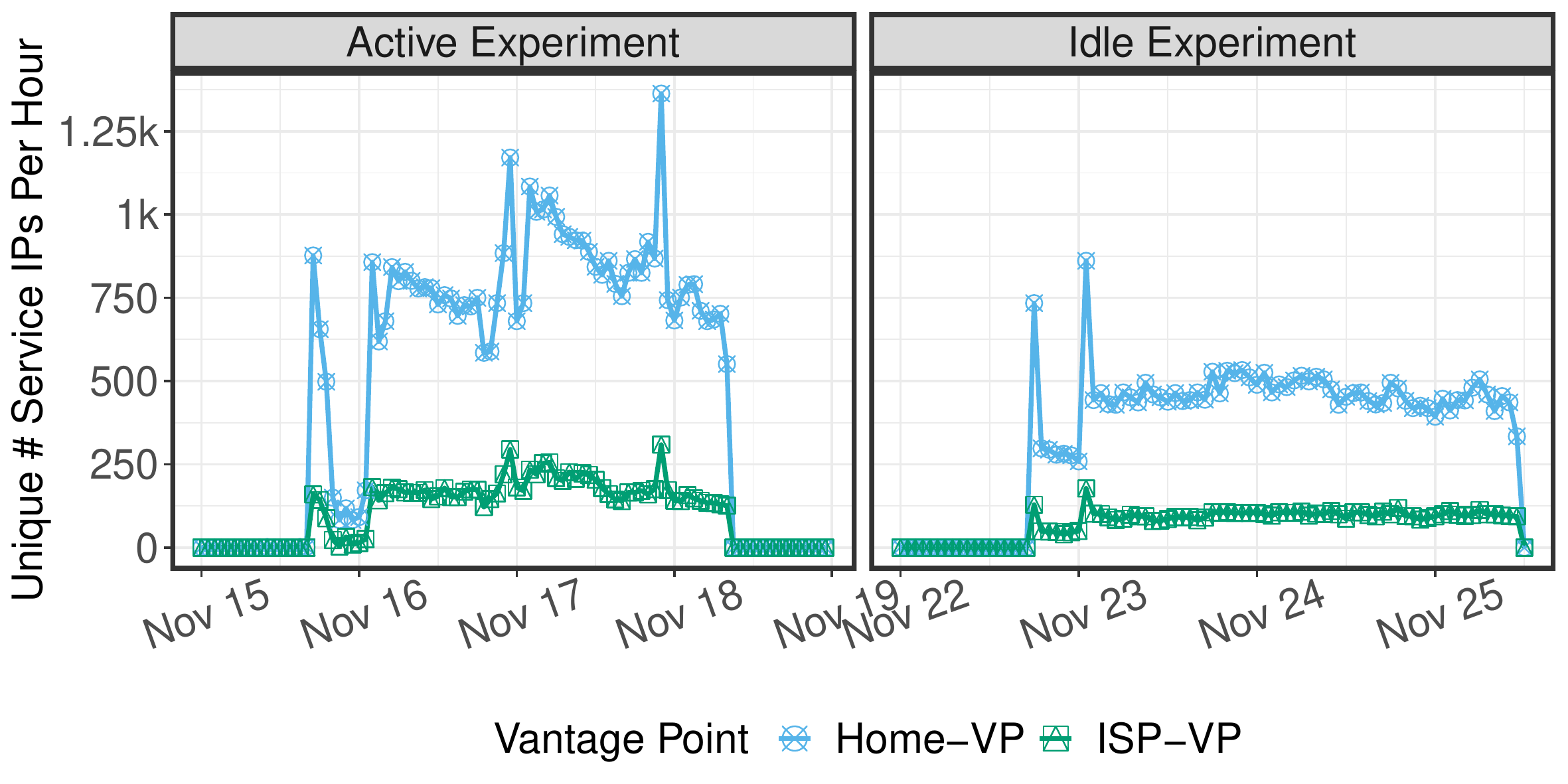}
     \label{fig:hourly-unq-dst-src-isp}
   }
   \subfigure[\# Unique domains per hour.]{
     \includegraphics[width=0.48\linewidth ]{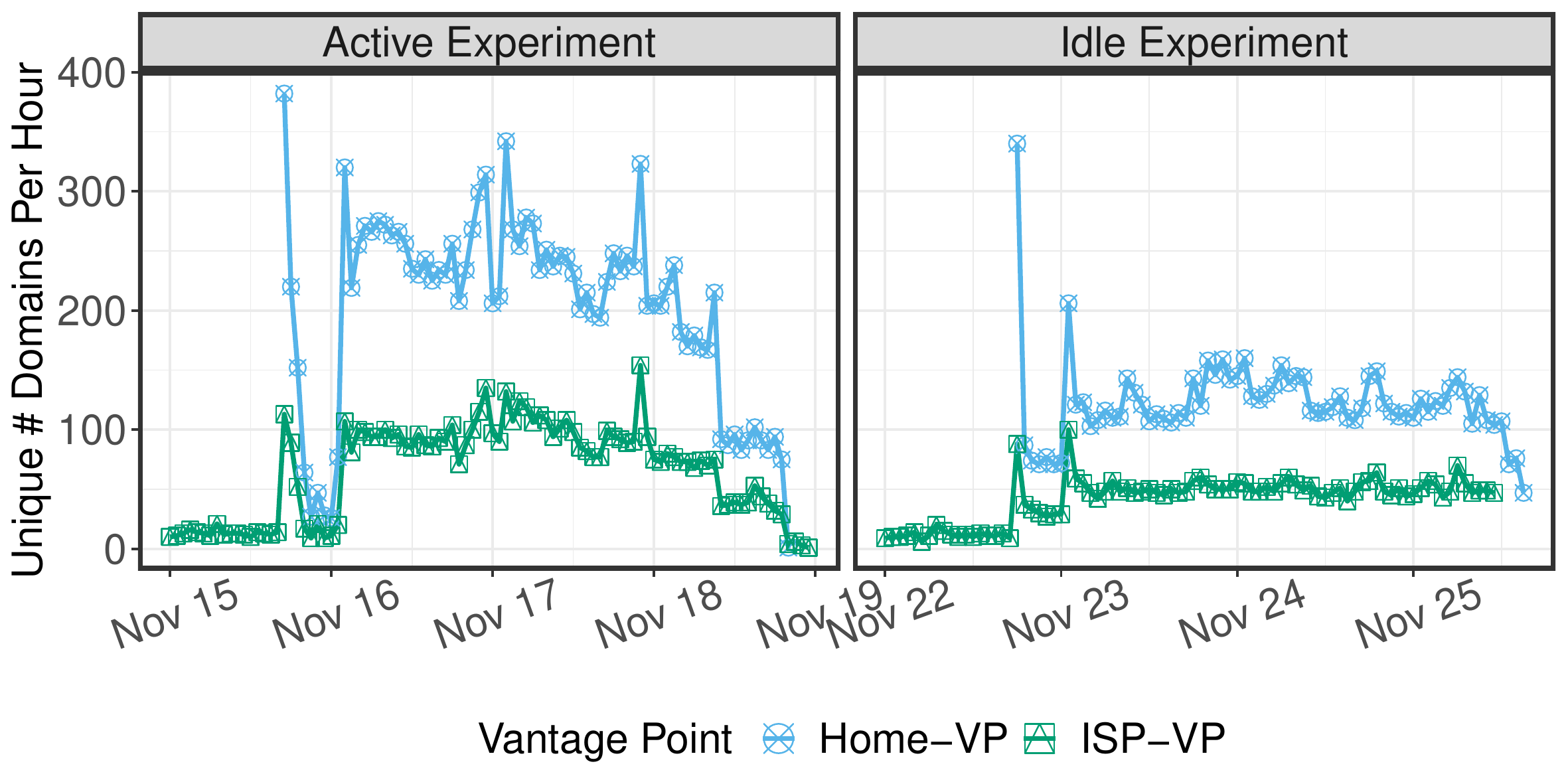}
     \label{fig:domain-visibility}
   }
 \end{minipage} 

 \begin{minipage}[t]{0.48\linewidth}
   \subfigure[Cumulative service IPs per port.]{
     \includegraphics[width=1\linewidth]{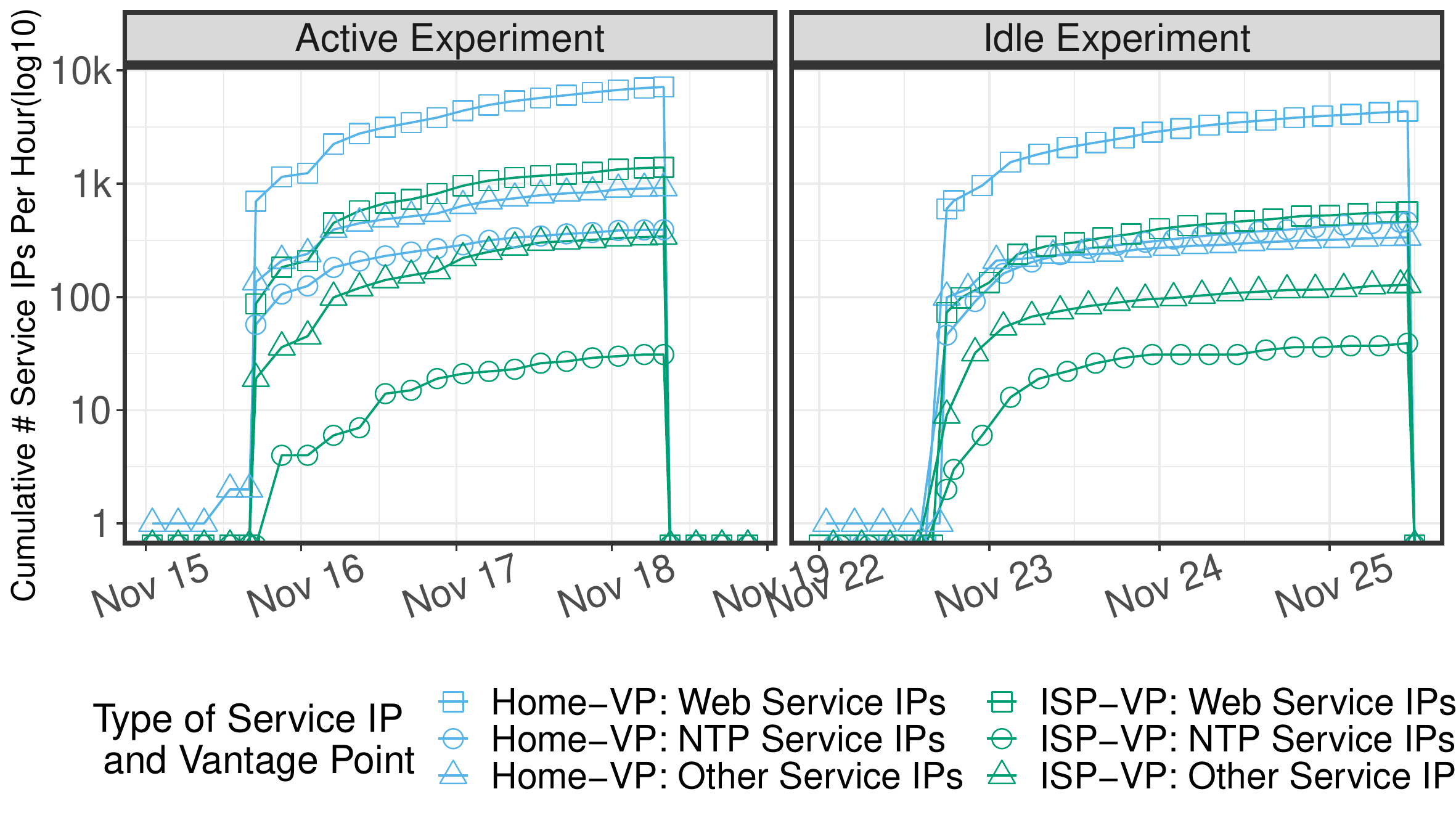}
     \label{fig:hourly-unq-dst-src-isp-cummulative}
   }
\end{minipage}
 \begin{minipage}[t]{0.48\linewidth}
	\subfigure[\# Unique IoT devices per hour.]{
		\includegraphics[width=1\linewidth]{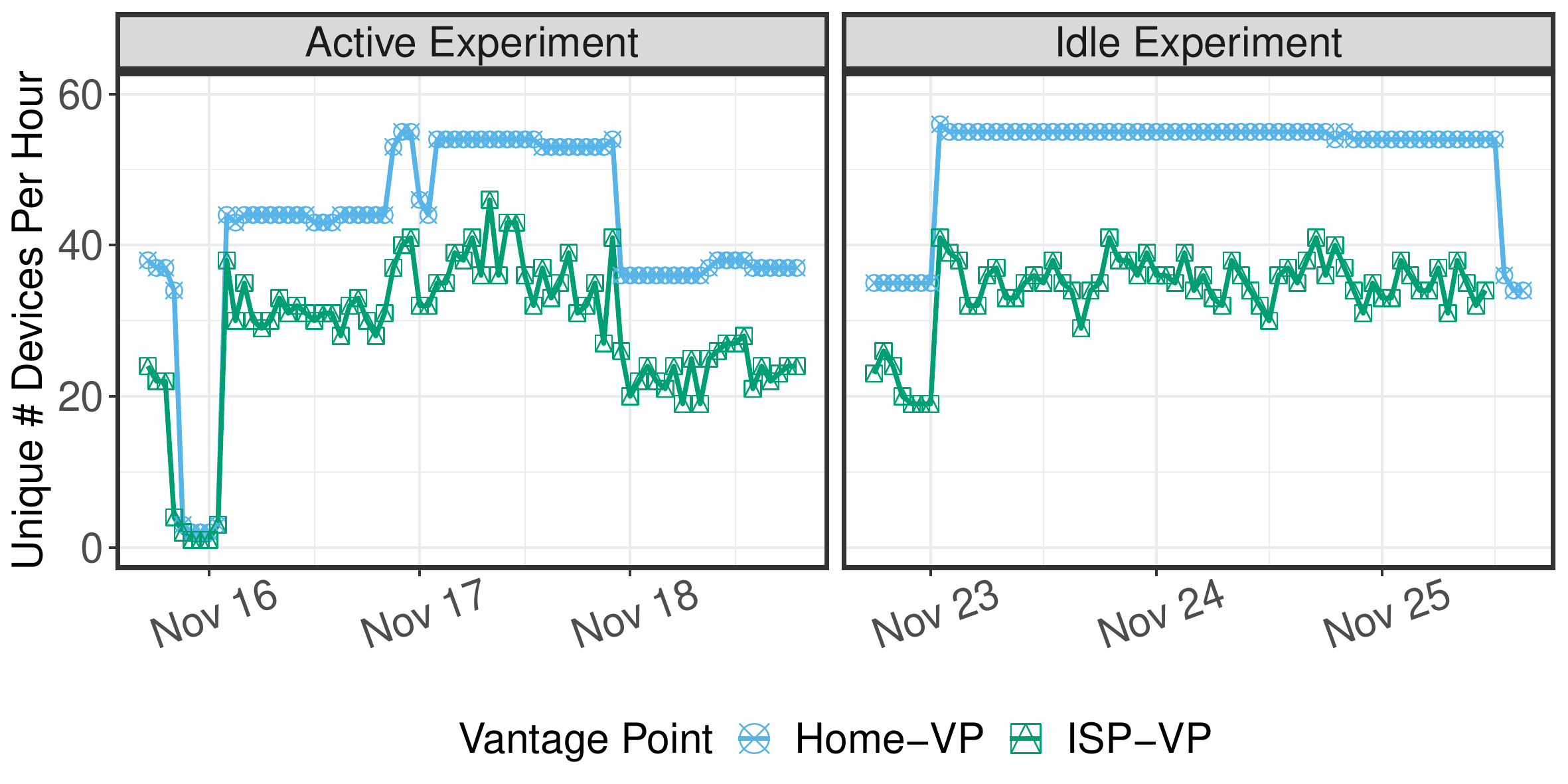}
		\label{fig:hourly-unq-dst-src-isp-devi}
	}

 \end{minipage}

   \captionsetup{skip=0.25em,font=small}
 \caption{Home-VP vs.\ ISP-VP.}
\end{figure*}
\begin{figure}[t!]
  \captionsetup{skip=0.25em,font=small}
 \begin{minipage}[t]{1\linewidth}
     \includegraphics[width=1.0\linewidth ]{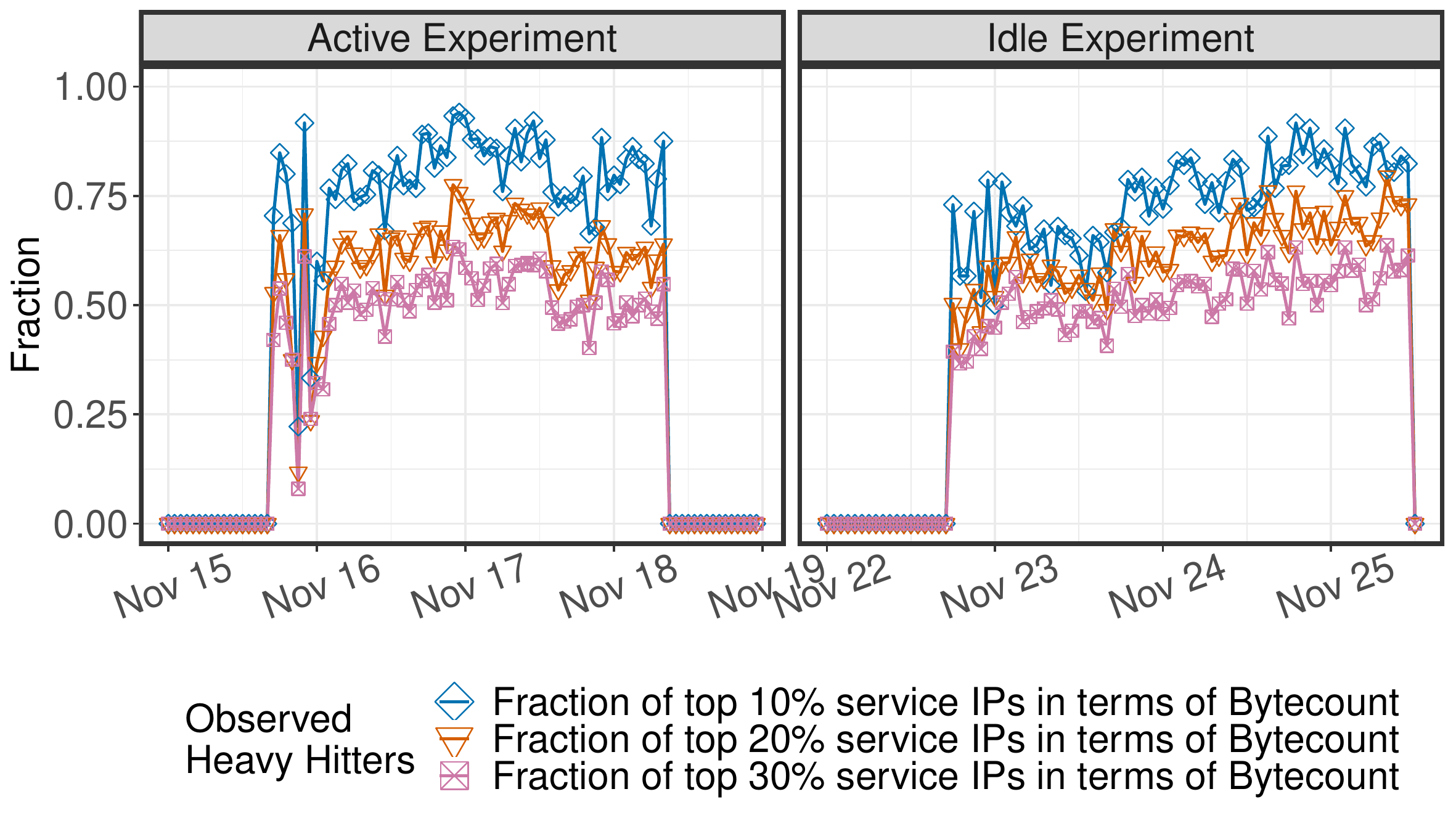}
  \captionsetup{skip=0.25em,font=small}
   \caption{ Fraction of observed ISP-VP vs.\ Home-VP per hour for popular servers
     (heavy hitters).}
           \label{fig:HH-visibility}
 \end{minipage}
\end{figure}

\subsection{Active and Idle IoT Experiments}
\label{sec:experiments}

Our experiments can be classified into \emph{idle} and \emph{active} experiments.

\noindent \textbf{Idle experiments}. We define as \emph{idle} the experiments during which the devices are just connected to the Internet without being actively used.  
We generate idle traffic for three days (November 23rd-25th, 2019) from both testbeds.  

\noindent \textbf{Active experiments}. We define as \emph{active} the experiments involving automated interactions.
We perform two types of automated interactions, each one repeated multiple times: (i) \emph{power interactions}, since in a previous study~\cite{Information-Exposure-IMC2019} it was reported that many IoT devices generate significant traffic when they are powered off and on. We manage the power status of the devices through several TP-Link smart plugs that we can control programmatically, followed by two minutes of traffic capture; (ii) \emph{functional interactions}, by automatically controlling the main functionality of the devices (\ie the act of switching on/off the light for a smart bulb) via voice (either directly or through a smart speaker) or via a companion app running on a separate network with respect to the IoT device (to force the communication to happen over the Internet rather than locally).
Unfortunately, some interactions for some devices cannot easily be automated (devices with \emph{idle} in Table~\ref{table:devices}). 
For these devices, we consider only idle experiments. In total, we perform~\numexperiments{} active experiments between November 15th and 18th, 2019.

\section{ IoT Traffic -- Visibility}\label{sec:visibility}

In this section, we aim to understand (i) to which extent the IoT related traffic of a \emph{single subscriber line} reaches a diverse set of servers in the Internet, and (ii) whether the low sampling rate of NetFlow limits the subscriber/device visibility. 
For this, we rely on the ground truth traffic for the Home-VP.  
More specifically, we monitor the IoT traffic at both vantage points: the Home-VP, as well as the border routers of the ISP-VP (see \circled{1} and \circled{2} of Figure~\ref{fig:general-overview}).

We first focus on the number of IP addresses that are contacted in each hour during the idle and the active experiments by the IoT devices, as stated in Section~\ref{sec:experiments}. 
We explicitly exclude DNS traffic, since it is not IoT-specific.
From Figure~\ref{fig:hourly-unq-dst-src-isp}, we see that during the active experiments, the IoT devices contact between 500 and 1,300 service IPs per hour when monitored at the Home-VP. 
Due to sampling, not all of this traffic is visible at the ISP-VP. 
We define \emph{service IPs} as the sets of IPs associated with the backend infrastructures that support the IoT services.
Indeed, the number of observed service IPs per hour in the ISP-VP decreases to an average of 16\%. 
Overall, during our idle experiments, the total number of contacted service IPs is lower, but the average percentage of observed service IPs remained at 16.5\%.

The spikes in the active experiments are partially due to power and the functional interactions. This can be seen on the idle experiments, where the spike indicates the action of starting the device (only at the beginning). 
Note that these spikes are also visible in the sampled ISP NetFlow data.

At first glance, 16\% sounds like a very small percentage. 
However, we note that the visibility of popular service IPs is significantly high.
Figure~\ref{fig:HH-visibility} shows the fraction of service IPs that are visible for the servers contacted the most, according to byte count. 
For the top 10\% of the service IPs, more than 75\% are visible, rising up to 90\% during some experiments. 
For less popular service IPs, e.g., the top 20\% and top 30\%, the visibility is only reduced to 70\% and 60\% in the active experiment, and a bit lower for the idle experiment. 

If we consider the entire period of our experiments, the percentage of visible service IPs is more than 34\% and 28\% for idle and active experiments.
Overall, at the daily level, more than 95\% of service IPs are visible for the top 20\%. 
Although we cannot observe \emph{all} IoT devices activity at the ISP-VP, a significant subset is visible.

While any specific service IP may not matter that much for an IoT service, its communications with a server domain name that may be hosted on multiple service IPs is essential. 
From the Home-VP, we know which service IPs correspond to which domain. 
Thus, we can determine which observed service IPs at the ISP-VP belong to which domain. 
This information is relevant for our methodology because in the ISP NetFlow data only IPs are visible. 
Figure~\ref{fig:domain-visibility} shows the number of observed \new{Fully Qulified Domain Names (FQDNs, we will refer to them as domains or domain names for the rest of the paper)} at the Home-VP and the ISP-VP. 
Many domains are hosted at multiple service IPs, hence we see that the number of observed service IPs is higher than the number of observed domains. 

\new{Figure~\ref{fig:hourly-unq-dst-src-isp-devi} shows the number of observed IoT devices per hour from the ground truth IoT traffic.  We observe a device when at least one packet from that device is seen within an hour. Note, For active mode, the experiments on devices from Testbed 1 (see figure \ref{fig:ISP-measurement}),  are initiated after Testbed 2. Therefore, all devices are not active during the same period. The average percentages of devices visible at ISP-VP, during active and idle experiments are 67\% and 64\% respectively.}

Next, we separate the observable network activity by ports. 
More specifically, we consider Web Services (ports 443, 80, 8080), NTP services (port 123), and other services (the rest of the ports), and we show the cumulative number of service IPs contacted. 
The resulting plot, Figure~\ref{fig:hourly-unq-dst-src-isp-cummulative}, shows that (i) the trend of observable service IPs at the Home-VP is mirrored at the ISP-VP, even when different services are considered, and (ii) the number of service IPs converges over time.

We also checked if any of the traffic from the Home-VP is visible at the IXP. 
However, neither during the active, nor during the idle experiments, we observe traffic at the IXP. 
This is expected as the ISP is not a member of the IXP.
Rather it peers directly (via private interconnects) with a large number of content and cloud providers as well as other networks. 

In summary, our analysis of the ground truth IoT traffic shows that, despite the low sampling of NetFlow, popular domains, service IPs, and ports of a {\it single} subscriber line (the Home-VP) are visible at the ISP.

\section{IoT Device detection methodology}\label{sec:methodology}

In this section, we outline our methodology for the detection of IoT devices in-the-wild. 
IoT services typically rely on a backend support infrastructure (see Figure~\ref{fig:iot-communication-pattern}) for user interactions.
From our ground truth experiments, we noticed that this backend infrastructure is often also used for keep-alives, heartbeats, updates, maintenance, storage, and synchronization. 
This observation is consistent with previous works~\cite{Information-Exposure-IMC2019, CIOTDI20_MazharS20}. 

We focus on identifying which Internet backend infrastructure is supporting \emph{each} of the IoT devices that we deployed in our testbeds (see \circled{3} in Figure~\ref{fig:general-overview}). 
When we refer to Internet backend infrastructure, we use two different abstractions: (i) sets of IP addresses/ports combinations as observable from the Internet vantage points, and (ii) sets of DNS domains. 
We focus also on domains because they are the primary indirect way for the devices to access their backend infrastructure.
While domain names are typically part of the permanent programming of the devices, IP addresses are discovered during DNS resolution, and may change over time.

A naive approach for identifying the backend infrastructure would be to use the ground truth traffic to identify which domains, and as a consequence, which service IPs are being contacted by each device. 
However this is not sufficient for the following reasons:

\begin{description}
\item [Limited relevance of some domains:] Not all domains are essential to support the services, or are useful for classification; for example, some domains may be used for advertisements or generic services,~\eg \texttt{time.microsoft.com} or \texttt{wikipedia.org}, see Section~\ref{sec:methodology-domains}.
\item [Limited visibility of IP addresses:] Since the ground truth data is captured at a single subscriber line only and DNS to IP mapping is rather dynamic, just looking at this traffic is not sufficient, see Section~\ref{sec:methodology-dnsdb}. 
\item [Usage of shared infrastructure:] Not all IoT services are supported by a dedicated backend infrastructure. 
Some rely on shared ones, such as CDNs. In the former case they can still have dedicated IP addresses; in the latter cases they use shared IP addresses, see Section~\ref{sec:methodology-dnsdb}. 
\item [Churn: ] DNS domain to IP address mappings are dynamic, see Section~\ref{sec:methodology-dnsdb}.  
\item [Common programming APIs: ] Multiple IoT services may use the same common programming API or may be used by different manufacturers; as a result, they often rely on the same infrastructure. This is the case for relatively generic IoT services such as Alexa voice service.
While this IoT service is available on dedicated devices, e.g., Amazon Echo, it can also be integrated into third-party hardware, e.g., fridges and alarm clocks~\cite{alexavoiceservice}. 
We cannot easily distinguish these from network traffic observations.  
\end{description}

\begin{figure}[t]
 \captionsetup{skip=.25em,font=small}
	\centering
  \includegraphics[width=1\linewidth]{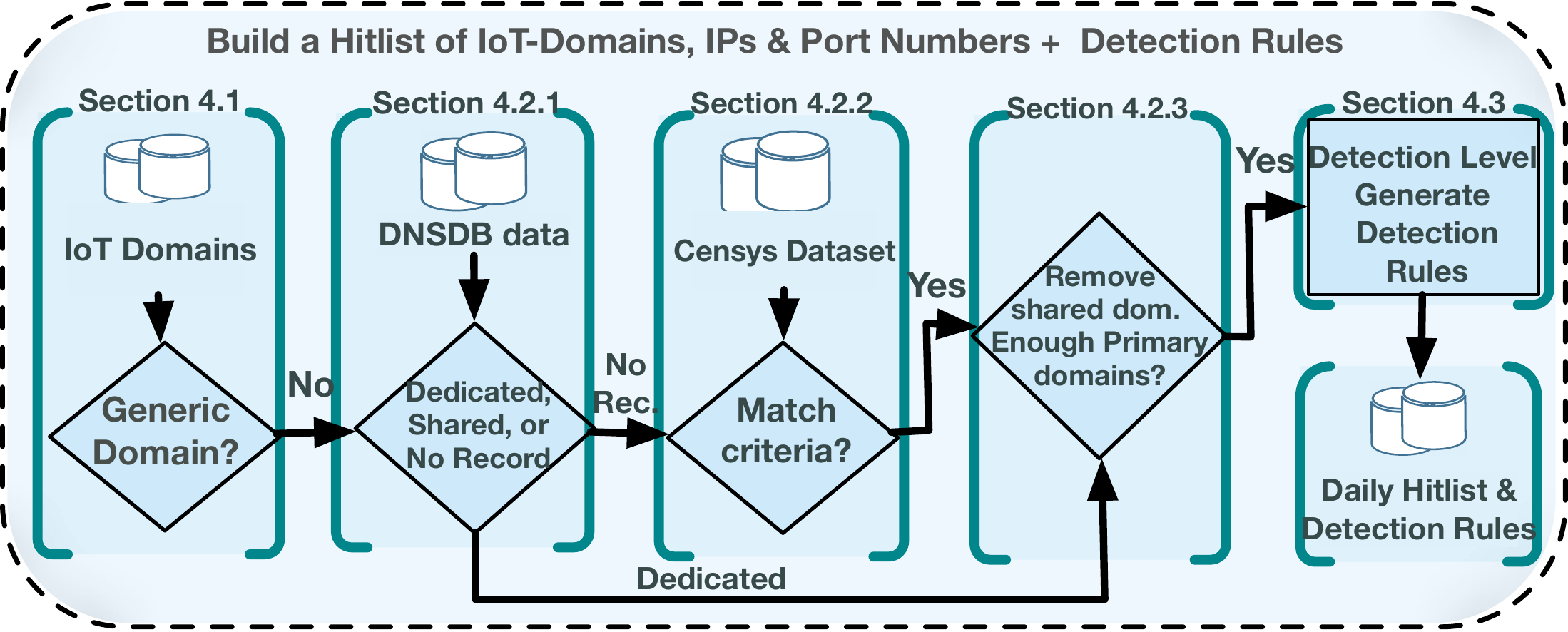}
\caption{IoT Traffic detection methodology overview.}
	\label{fig:methodology}
\end{figure}

\begin{figure}[t]
	\captionsetup{skip=0.25em,font=small}
	\centering
	\includegraphics[width=0.45\textwidth]{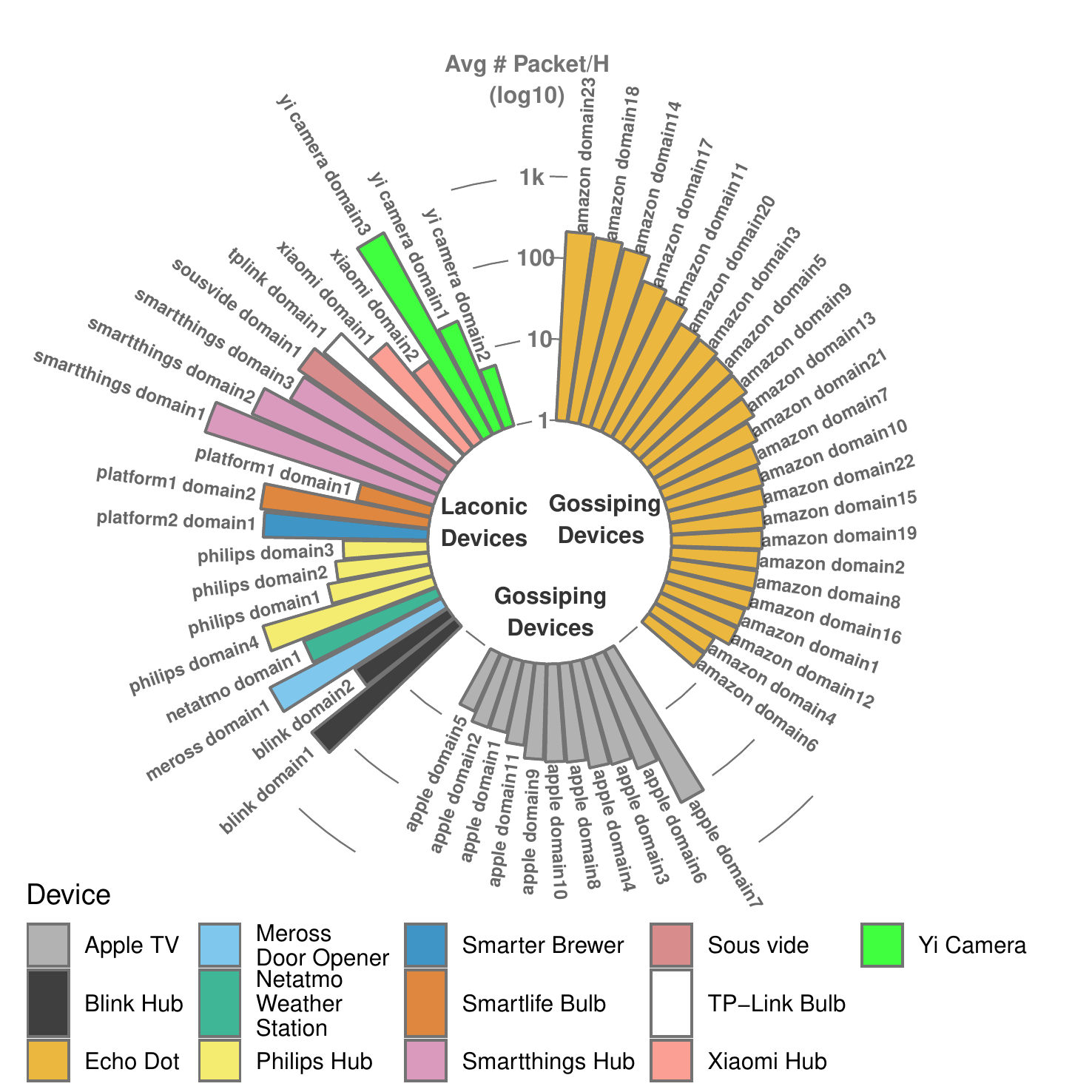}
	\caption{Home-VP: Circular bar plot of average \# of packets/hour  per domain (log y-scale). The domains belong to 13 IoT devices and separated into three groups: one for laconic and two for gossiping devices (Echo Dot and Apple TV).}
	\label{fig:vis-domain}
\end{figure}

Below we tackle these challenges one by one. 
The outcome is an IoT dictionary that contains mappings for individual IoT services to sets of domains, IP addresses, and ports. Based on IoT services, we generate rules for IoT device detection.  
For an overview of the resulting methodology, see Figure~\ref{fig:methodology}. 

\subsection{Classifying IoT Domains}\label{sec:methodology-domains} 

The amount and frequency of network traffic that an IoT device exchanges with its backend infrastructure varies from device to device, depending on the complexity of its services, its implementation specifics, and the usage of the device. 
This is highlighted in Figure~\ref{fig:vis-domain}, where we show the average number of packets per device and per domain (using a log y-scale) for $13$ different devices (subset of devices) in their idle mode. 
The first observation is that most devices are supported by their own set of domains and for many IoT services, this is a small set containing less than 10 domains. 
We refer to these as \emph{small domain sets} as they correspond to \emph{laconic} devices. 
Other devices \emph{gossip} and have \emph{sizable domain sets}.
Figure~\ref{fig:vis-domain} shows the domains of two example gossip devices (Apple TV in gray and Echo Dot in orange) and several laconic devices (rest of the colors).

Having a sizable domain set often indicates the usage of a larger infrastructure, which may not be dedicated to a specific IoT service.  
We find that most of these domains are mapped via CNAMEs to other domains. 
For the two gossiping examples considered in Figure~\ref{fig:vis-domain}, the domains of Echo Dot are mostly mapped to its own infrastructure. 
However, the ones of Apple TV are mainly mapped to a CDN---in this case, Akamai---that offers a variety of services.

Based on these observations from our ground truth data, we classify the domains as follows:

\begin{description}
\item [IoT-Specific domains.] Grouped into (i) \emph{Primary} domains: registered to an IoT device manufacturer or an IoT service operator; and (ii) \emph{Support} domains: that are not necessarily registered to IoT device manufacturers or service operators, but offering complementary services for IoT devices, \ie \emph{samsung-*.whisk.com} for Samsung Fridges, here whisk.com is a service that provides food recipes and images of food. 
\item [Generic domains.] Domains registered to generic service providers that are heavily used by non-IoT devices as well, e.g., \emph{netflix.com}, \emph{wikipedia.org}, and public NTP servers.
\end{description}

\begin{figure}[t]
	\captionsetup{font=small}
	\centering
	\includegraphics[width=0.8\linewidth]{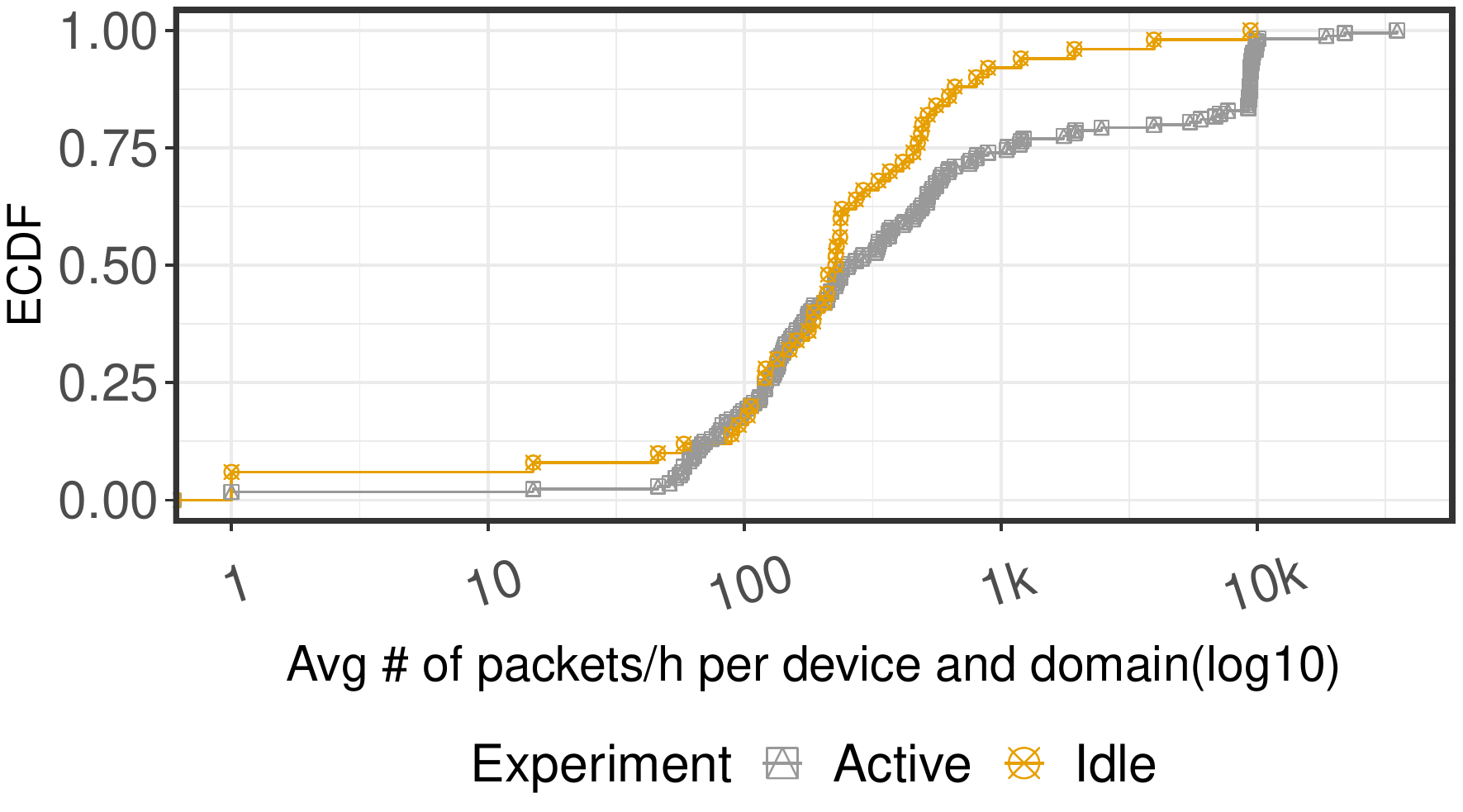}
	\caption{Home-VP: ECDF of average \# of packets/hour for all IoT-Specific domains, per device, (idle and active experiments).}
	\label{fig:active_idle}
\end{figure}

We classify each domain name from our idle and active experiments using pattern matching, manual inspection, and by visiting their websites and those of the device manufacturers. 
Since the \emph{Generic} domains cover non-IoT traffic, we do not further consider them. 
Rather, we focus on the  \emph{IoT-Specific} domains. 
As a result, we classify 415 out of the 524 domains as \emph{Primary} and 19 as \emph{Support} domains. 

Next, we explore the volume of traffic that the IoT devices exchange with all domains. 
Figure~\ref{fig:active_idle} shows the ECDF of the average number of packets per hour per domain for all IoT-Specific domains for both the idle and the active experiments.
First, we note that almost all devices and domains \new{, except for one device in its idle mode,} are exchanging at least 100 packets per hour, and this \emph{may not suffice for detecting} them in any given hour in the wild due to sampling. 
However, during the active experiments, we see that some domains are only used when the device is \emph{active} or other domains receive significantly more traffic, up to and exceeding 10K packets, which \emph{may suffice for detection}. 
These latter domains may be ideal candidates for detecting such devices in the wild.

\subsection{Identifying Dedicated Infrastructures}

Once we have a list of IoT-Specific domains (FQDNs) with their associated service IP addresses and port mappings from the ground
truth experiments, we need to understand whether they have a shared or dedicated backend infrastructure. The reason is
that, if we want to identify IoT services and consequently IoT devices in the wild by using network traces such as
NetFlow, we can only observe standard network level features such as src/dst IP and port numbers without packet payload.
Therefore, if a service IP belongs to a shared infrastructure such as a CDN or a generic web hosting service, this
service IP can serve many domains, and it is impossible for us to exactly know which domain was actually contacted. To
this end, the purpose of this section is two-fold. First, to expand the candidate service IPs beyond those directly
observed in the ground truth experiments (to mitigate that we are focusing on a single subscriber line). Second, to
classify domains into those that use backend services hosted on dedicated infrastructure service IPs vs.\ those that
rely on shared infrastructure service IPs. We do this by relying on DNSDB~\cite{dnsdb},
Censys~\cite{CCS15_DurumericAMBH15}, and applying additional filters.

\subsubsection{From IoT-Specific Domains to Service IPs: DNSDB}\label{sec:methodology-dnsdb}

We use IoT-Specific domains to identify the backend infrastructure that is hosting them. 
To this end, we leverage the technique in~\cite{IMC18_IordanouSIL18}, and use these domain names to identify all associated service IPs on which these domains are hosted during the time period of our experiments. 
We use both the ground truth experiments, and external DNS databases, including DNSDB~\cite{pDNS:2005}. 
We found that the specific IP addresses mapping to specific domains can change often.  
However, DNSDB provides information for all domains served by an IP address in a given time period and vice versa, hence it mitigates the issues caused by this churn.
DNSDB also provides all records, including CNAMEs that may have been returned in the DNS response, for a given domain. 
Thus, we use DNSDB to check if a service IP address is \emph{exclusively} used for a specific IoT service, or if it hosts additional domains.  
We say a service IP is \emph{exclusively} used if it only serves domains from a single ``second-level'' domain (SLD) and its CNAMEs.
However, we note that the CNAMEs may not involve the same second-level domain. 
Let us consider an example: the domain \texttt{devA.com} is mapped via a chain of CNAMEs
such as \texttt{devA-VM.ec2compute.amazonaws.com} to IP \texttt{a.b.c.d}. This IP only reverse maps to
\texttt{devA-VM.ec2compute.amazonaws.com} and its associated CNAME
\texttt{devA.com}. Since this is the only CNAME associated with the IP, we
may consider this IP a direct mapping for the domain. Yet, at the same time, we
find support that public IP addresses assigned to a cloud resource such as a virtual machine in AWS EC2, that is occupied by a tenant, is not shared with other tenants unless the current resource is released. This is a popular service offered by multiple platforms~\cite{awsvpc,awselasticcomputeip,msazureip}.
Let us consider a second example: domain \texttt{devB.com}. It may use the
Akamai CDN. Thus, the domain \texttt{devB.com} is a CNAME for
\texttt{devB.com.akadns.net}. This domain then maps to IP
\texttt{a.b.c.d}. However, in this case, many other domains, e.g.,
\texttt{anothersite.com.akadns.net}, also map to this IP. Thus, we may conclude that
this domain is hosted on a \emph{shared} infrastructure.

Once we understand if an IP is exclusively used for a specific IoT service, we can also classify the domains as either using a \emph{dedicated} or \emph{shared} infrastructure. 
For the former, all service IPs have to be dedicated to this domain for all days, otherwise we presume that the domain relies on a shared infrastructure.

Once we apply this methodology to all 434 domain names, we find that 217 are
hosted on dedicated service IPs, while 202 are relying on a shared backend
infrastructure. For 15 of the domains we did not have sufficient information in
DNSDB. We handle them in the next step.

\subsubsection{From IoT-Specific Domains to Service IPs: Censys}\label{sssec:censys}

Among the reasons that DNSDB may not suffice for mapping some domains to service IPs is that
(a) frequent remapping of domains to IPs or, (b) missing data since the requests
for the domains may not have been recorded by DNSDB, which intercepts requests
for a subset of the DNS hierarchy. To overcome this limitation, we rely on the
certificate and banner datasets from Censys~\cite{CCS15_DurumericAMBH15}, to infer the ownership
of the domains and the corresponding IPs, as long as these are using HTTPS. 
For example, we did not find any record for the domain \texttt{c.devE.com} in the DNSDB dataset. We then check if device $E$ uses {\tt HTTPS} to communicate with this domain. This allows us to query for all service IPs that potentially offer the same web certificate as the hosts in this domain.
For a certificate to be associated with a domain, we require that the domain
name and the \emph{Name} field entry in the certificate match at least the SLD or higher, i.e. the Name field of the certificates matches the pattern \texttt{c.devE.com} or \texttt{*.devE.com}
and that there is no other \emph{Subject Alternative Name (SAN)} in the certificate.
Next, we query the Censys dataset for all IPs with the same
certificate and {\tt HTTPS} banner checksum for the domain from our ground truth dataset within the same period.  
This allows us to identify data for 8 out of 15 of the domains which belong to 5 devices. 

\subsubsection{Removal of Shared IoT Backend Infrastructures}\label{sec:removal-shared}

In the last step of our methodology we filter out devices that use shared backend infrastructures.
We find that Google Home, Google Home Mini, Apple TV, and Lefun camera, all have a shared backend infrastructure. 
For LG TV, we are left with only one out of 4 domains; for Wemo Plug and Wink-hub, we could not identify sufficient information.
Because of this, we have excluded these devices from further consideration.

The result forms our \emph{daily} list of dedicated IoT services, along with their associated domains, service IPs and port combinations.

\subsection{IoT Services to Device Detection Rules}\label{sec:detection-rules}

Once we identified the set of IoT services that can be monitored, we generate the rules for detecting IoT devices. Depending on the set of IoT services contacted by the devices we can generate device detection rules at three granularity levels: (i) Platform-level,  (ii) Manufacturer-level, and (iii)
Product-level, from the most coarse-grained to the most fine-grained, respectively.  In this section, first, we show how
we determine the detection level for each device.  Then, we explain how we generate the detection rules for each IoT
device for the detection level that can be supported.

\subsubsection{Determining IoT Detection Level}
\begin{description}
\item [Platform-level:]Some manufacturers use off-the-shelf firm\-ware, or outsource their backend infrastructure to IoT platform solution companies such as Tuya~\cite{tuyaplatform}, electricimp~\cite{electricimp}, AWS IoT Platform~\cite{awsiot}.
These IoT platforms can have several customers/manufacturers that rely on their infrastructure. 
Therefore, we may not be able to distinguish between different manufacturers from their network traffic.  

\item [Manufacturer-level:] The majority of our studied IoT services rely on dedicated backend infrastructures that are
operated by the manufacturers themselves.  We also observe that many manufacturers rely on similar APIs and backend
infrastructures to support their different products and services.  This makes distinguishing individual IoT products
from their network traffic more challenging. 

\item[Product-level:] This is the most fine-grained detection level, where we are able to distinguish between different
products of a manufacturer, e.g., Samsung TV, or Amazon Echo vs. Amazon Fire TV.  For detection at the product level, we underline
the importance of side information about the purpose associated with a domain.  With this information, we can improve
our classification accuracy. For example, for Alexa Enabled devices, the domain \texttt{avs-alexa.*.amazon.com} is
critical, as it is the base URL for the Alexa Voice Service API~\cite{alexavoiceservice} (shown in
Figure~\ref{fig:vis-domain} as amazon domain23).  Other examples are the Samsung devices that use the domain
\texttt{samsungotn.net} to check for firmware updates~\cite{samsung-firmware-update-servers}.

\end{description}

Additionally, some advanced services of the devices often require additional backend support from manufacturers. 
These may then contact \emph{additional} domains.  
By considering more specific features (domains), the capabilities to distinguish products increases.  
We leverage these specialized features \eg to distinguish Amazon Fire TV, which contacts significantly more domains than other Amazon products, \eg Echo Dot. 

\subsubsection{Generation of Detection Rules}
For any of our three levels of detection, 
we require that a subscriber contacts at least one IP/port combination associated with a Primary domain of the IoT service, to claim detectability of IoT activity at the subscriber. 
However, if there are many domains, requiring only one such activity may not have enough evidence. For example, by monitoring a single domain we can detect all Alexa Enabled devices, but this service can be integrated into third party hardware as well. Therefore, in order to detect products manufactured by Amazon, e.g., Amazon Echo, it is essential to monitor additional domains that are contacted by the Amazon Echo devices.     
For this, we introduce the \emph{detection threshold} $D$. 
If an IoT service has $N$ IoT-Specific domains, we require to observe traffic involving $k$ IP/port combinations that are associated with $max(1, \lfloor {D}\times{N} \rfloor)$ of the $N$ domains. 
To determine an appropriate value for this threshold, we rely on our ground truth dataset, see Section~\ref{sec:crosscheck}.

We start with \numdevices devices in our testbeds. 
We have multiple copies of a same device deployed in different continents. 
This reduces the set of devices to \numproducts unique products. 
Of these, many are from the same manufacturer, \eg a Xiaomi rice cooker, a Xiaomi plug, and a Xiaomi light bulb. 
Since these devices are often supported by the same backend infrastructure of the manufacturer, the list of domains has significant overlap and often fully overlaps. 
In our methodology we can detect 3 different IoT platforms, the coarsest level, as 4 of our products rely on them. Moreover, we generated rules for the detection of 29 IoT devices at the manufacturer level. We had a diverse range of products from Amazon and Samsung in our testbed that allowed us an in-depth analysis, and cross-examination of domains contacted by different products. Therefore, for devices using Alexa voice service (\ie Alexa Enabled), and for Samsung IoT devices, we detect the former at the platform level and the latter at the manufacturer level. 
For Alexa Enabled and Samsung IoT devices, we compared the domains across different devices and obtained enough side information about the purpose of their domains that allowed us to further divide each of them into two subclasses at more fine grained levels.
For this, we defined a hierarchy, namely Amazon products, and Fire TV, under Alexa Enabled devices. Amazon products are detected at manufacturer level, and include products such as Amazon Echo family and is superclass of Fire TV. 
We identified 33 additional domains, besides the Alexa voice service domain, that were contacted by Amazon products.  
Moreover, Fire TV contacts up to 67 domains (34 more domains than Amazon products). This allows us to establish its subclass, at product level, under Amazon products. 
Using side information~\cite{samsung-firmware-update-servers} and comparing the set of domains across different Samsung products, we monitor 14 domains in total, but only one domain is important to detect Samsung IoT devices with Samsung firmware (these include a broad range of products, such as fridges, washing machines and TVs). Samsung TVs contact 16 additional domains that are not used by any of the other Samsung devices in our testbed.

\begin{figure} [t!]
	\captionsetup{skip=.5em,font=small}
	\centering
	\includegraphics[width=1\linewidth]{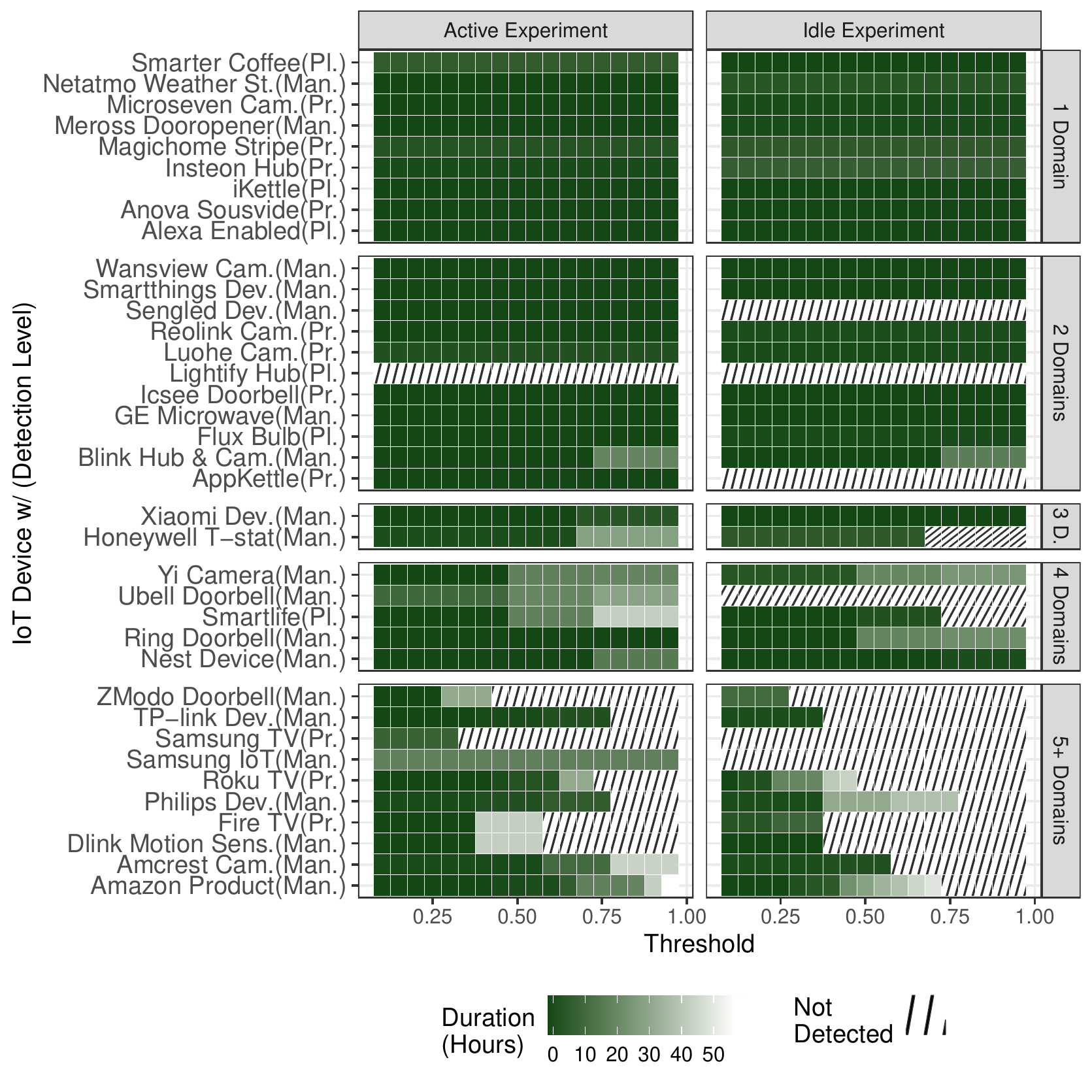}	
	\caption{Home-VP: Time to detect IoT (per threshold).}
	\label{fig:detection-delays}
\end{figure}

Using the above methodology, \new{ except for the devices listed in section~\ref{sec:removal-shared}, we generated detections rules at different levels for our testbed devices.} We generated rules for the detection of 20 manufacturers, and 11 products that amounts to the 77\% of manufacturers in our testbeds. We generate rules for 4 \emph{unique} IoT platforms by monitoring 1 to 4 domains (2 platforms were contacted by 4 devices, we report them separately). Finally, for 11 products we consider between 1 to 67 domains. For a detailed number of domains per IoT device see~Figure\ref{fig:detection-delays}.

\section{Methodology: Crosscheck}\label{sec:crosscheck}

We use our ground truth dataset to check how long it takes for our methodology (applied to the sampled flow data from the ISP) to detect the presence of the IoT devices for the idle and the active experiments (see \circled{4} of Figure~\ref{fig:general-overview}). 
For this, we report the time that it takes to detect an IoT device that is hosted in our ground truth subscriber line when it is in active mode (Figure~\ref{fig:detection-delays} left) and idle mode (Figure~\ref{fig:detection-delays} right).  We only include the ones that are detectable with our methodology, \ie those that do not rely exclusively on shared  infrastructures. We also annotate the device name with its detection levels: Platform (Pl.), Manufacturer (Man.), and Product level (Pr.). 
\begin{figure*} [t]
	\captionsetup{skip=.5em, width=0.75\linewidth,font=small}
	\centering
	\subfigure[Per Hour.]{     
		\includegraphics[width=0.48\linewidth]{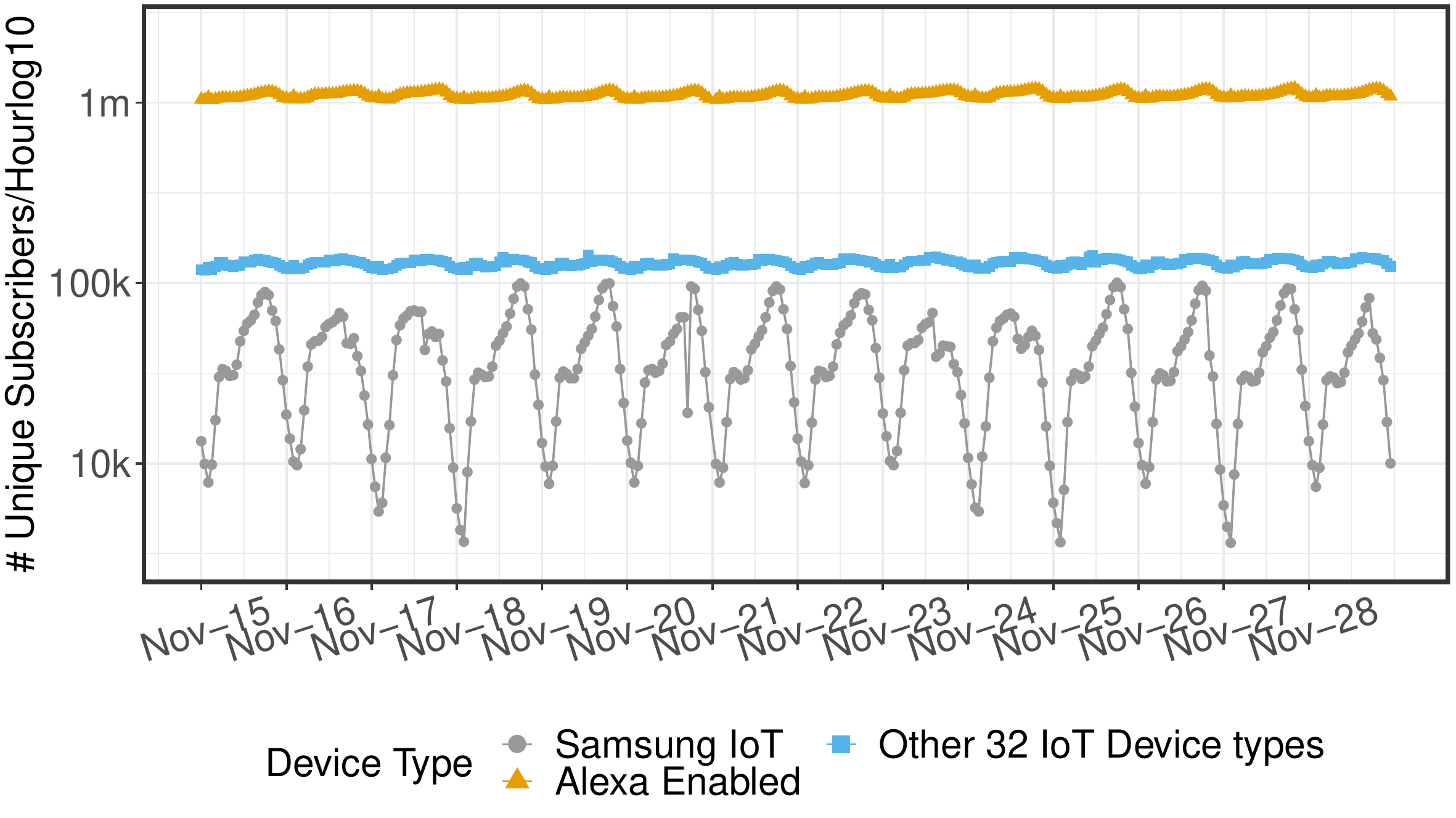}
		\label{fig:alexa-samsung-wild-dev-hour}
	}
	\subfigure[Per Day.]{        
		\includegraphics[width=0.48\linewidth]{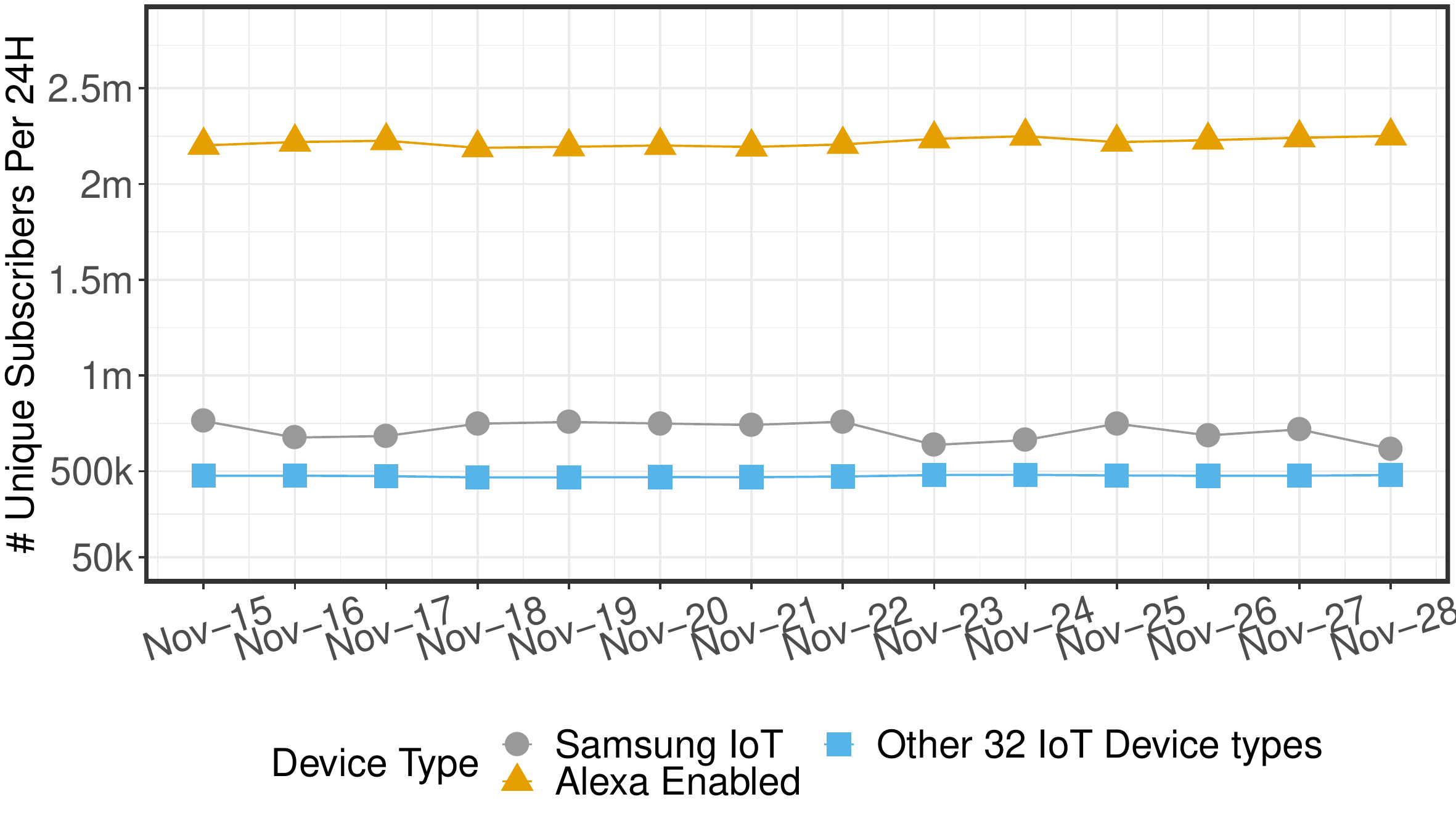}
		\label{fig:alexa-samsung-34dev-wild-dev-day}
	}
	\caption{ISP: Per Hour,  Subscriber lines with IoT activity (Alexa Enabled, Samsung
		IoT, and others).
		\label{fig:alexa-wild-dev-hour-day}}
\end{figure*}
\begin{figure}[t]
	\captionsetup{skip=.25em, width=1\linewidth,font=small}
	\includegraphics[width=1\linewidth]{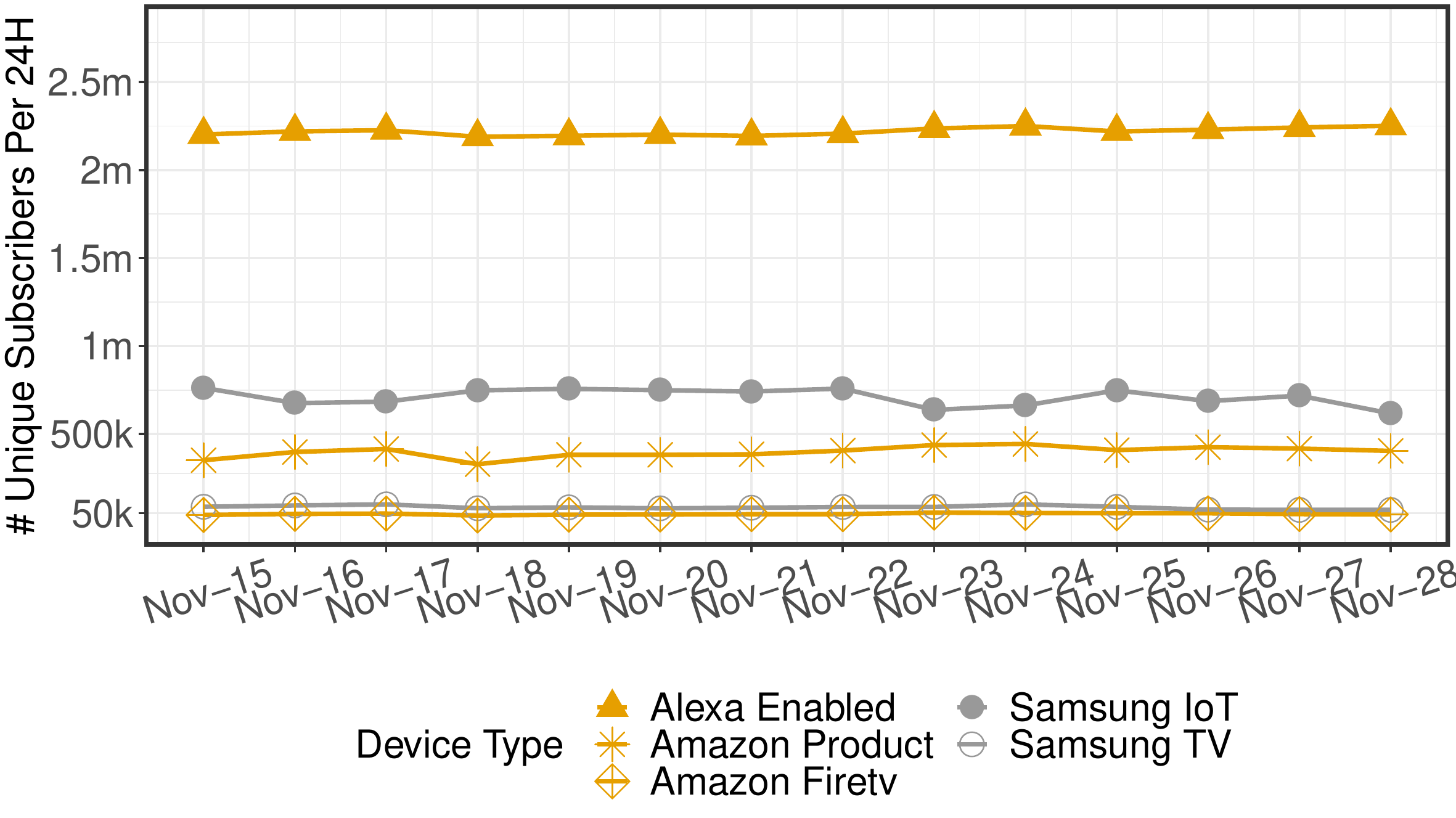}
	\caption{ISP: Drill down for Amazon and Samsung IoT devices--per  day.}
	\label{fig:alexa-sam-detailed-day}
\end{figure}

On average, by requiring the evidence of at least 40\% of domains, we are able to detect 72/93/96\% of IoT devices that are detectable at manufacturer or product level within 1/24/72 hours in the active mode. Even in idle mode their the percentage is 40/73/76\% with 1/24/72 hours. For the devices detectable only at product level (Pr.), with the same required evidence, we detected 63/81/90\% of them within the 1/24/72 hours respectively, in active mode. \new{Note, we are using the sampled ISP data.}
Indeed, popular products such as Amazon products (\ie Echo Dot, Echo Spot) can be almost instantly detected. 
This is a significant finding and underlines that it is possible to use sampled flow data within an ISP to accurately detect the
presence of a specific IoT product within a subscriber line, despite differences in
activity and IP churn due to operational requirements.

A closer look reveals that, in general, it takes longer to detect an idle IoT device in comparison to when it is active. 
This is not surprising, as most IoT devices show more network activity in active mode. 
However, this does not mean that the increase will occur across all of the services contacted by a device, since 
there are exceptions that take longer to detect even in active mode, e.g., SmartLife, and Nest.

Figure~\ref{fig:detection-delays} also contains information regarding the number of monitored domains per IoT device with their detection level. 
For 9 IoT devices, a single domain is considered. 
For the others, we consider many more (up to 67). A \emph{threshold} determines the fraction of domains for which we require evidence of network traffic to claim detection. 
To understand the impact of such threshold on detection time, we variate its value from 0.1 to 1 and show the corresponding detection times. 
Note, for IoT devices where we consider only one domain, the variation of the threshold does not change the detection time, as we always require evidence of at least one domain.
Overall, we note that a larger threshold can increase the detection time, and some IoT devices may no longer be detectable. 
However, it may also increase the false positive rate.
We crosscheck possible false positives by running another experiment where we only enable a small subset of IoT devices. 
We then apply our detection methodology to these traces and do not identify any devices that are not explicitly part of the experiment. 
We also try to avoid false positives by ensuring that the domain sets per device differ.

Regarding detectability, we notice that 6 IoT devices could not be detected even after the entire duration of our idle experiments.
A closer investigation shows that for 5 of these, the frequency of traffic is so small that their likelihood of detection is very low. 
Indeed, for this specific time period, they were invisible in the NetFlow data. \new{This highlights that in order to be able to confidently detect a device, the device have to either exchange enough packets with the targeted domains or the sampling rate shall be increased.} For Samsung TV, we require to observe enough domains to confirm the presence of a Samsung IoT device, before moving forward with detection. Thus, if we do not see enough Samsung IoT domains, then we do not claim the detection of Samsung TVs.
Nevertheless, the results look very promising for us to attempt on detecting deployed IoT devices in the wild. 


\section{Results: IoT in the Wild}\label{sec:iot-wild-results}

In this section, we apply our methodology for detecting IoT activity in the ISP and IXP data (see \circled{5} in Figure~\ref{fig:general-overview}).
For this we focus on the two weeks in which we collected the data from the ground truth experiments to obtain up-to-date mappings of domains to IPs.

\subsection{Ethical Considerations and Privacy Implications}

Applying our methodology to traffic data from ISPs and IXPs may raise ethical concerns as it may be considered as analyzing customer activities. 
However, this is not the goal of this paper. 
The goal here is to showcase that it is possible to detect and map the penetration of IoT device usage. 
As such, this study is not about subscribers' device activities, instead it is about detection capabilities and aggregated usage. 
Thus, we report on percentages of subscriber lines where we can observe IoT related activity. 
Indeed, we are unable to trace IoT activity back to individuals as the raw data was anonymized as per recommendations
by~\cite{IMC19_DekovenRMABSSVS19} and never left our collaborators' premises. \new{Moreover, we do not analyze any data
that is not related to the detection of IoT presence, e.g., DNS queries~\cite{iotfinder}, or flows that are not related to IoT
backend infrastructures, to eliminate any user Web visit profiling.}

\begin{figure} [t]	
	\captionsetup{skip=.25em,font=small}
	\includegraphics[width=1\linewidth,valign=t]{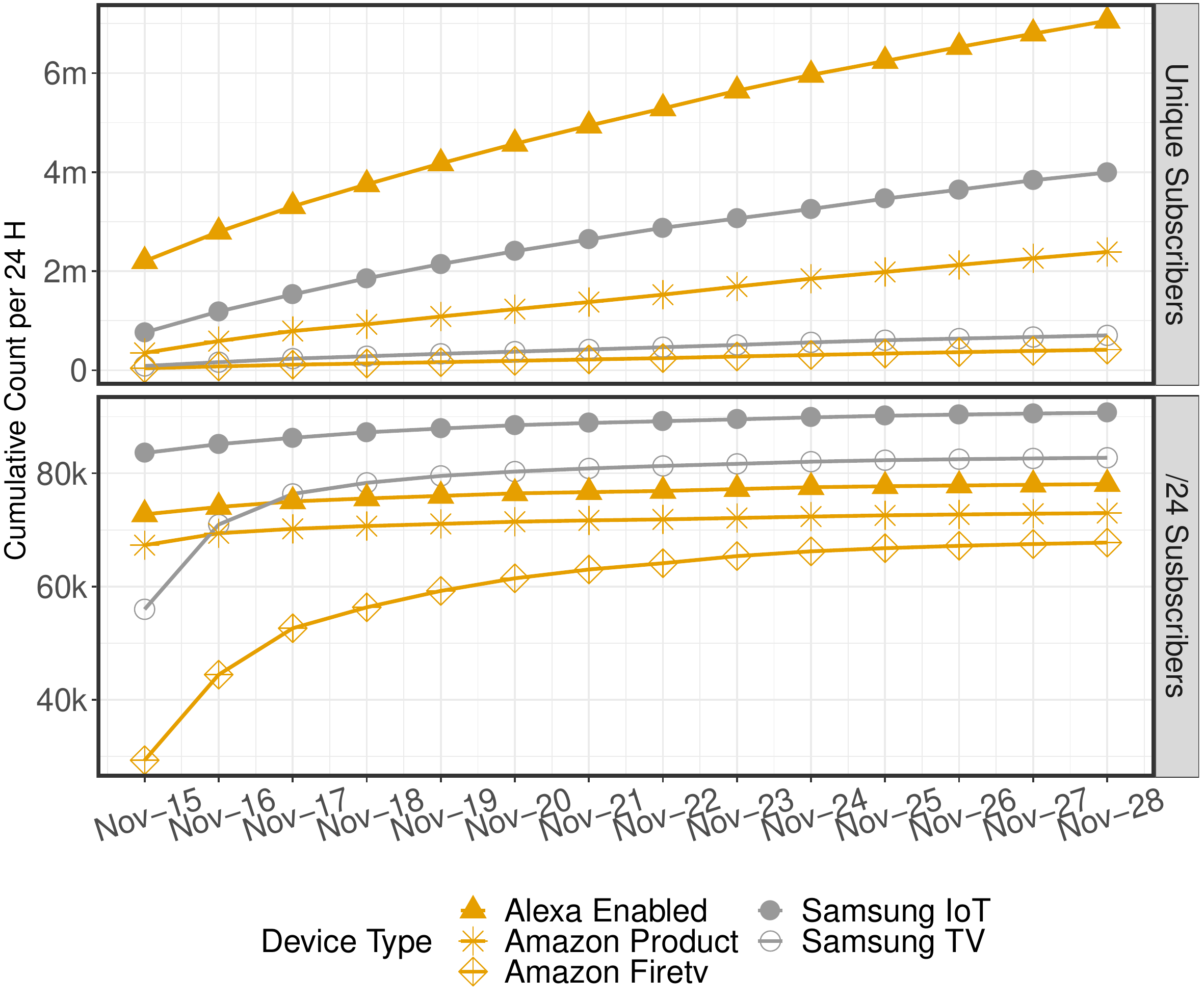}
	\caption{ISP: Cumulative \# of subscriber lines resp.\ /24s with daily IoT
		activity across two weeks.}
	\label{fig:alexa-wild-cumulative-day}
\end{figure}

\subsection{Vantage Point: ISP}\label{sec:isp-results}

\begin{figure}
	\captionsetup{skip=.4em,font=small}
	\centering
	\includegraphics[width=1\linewidth,valign=t]{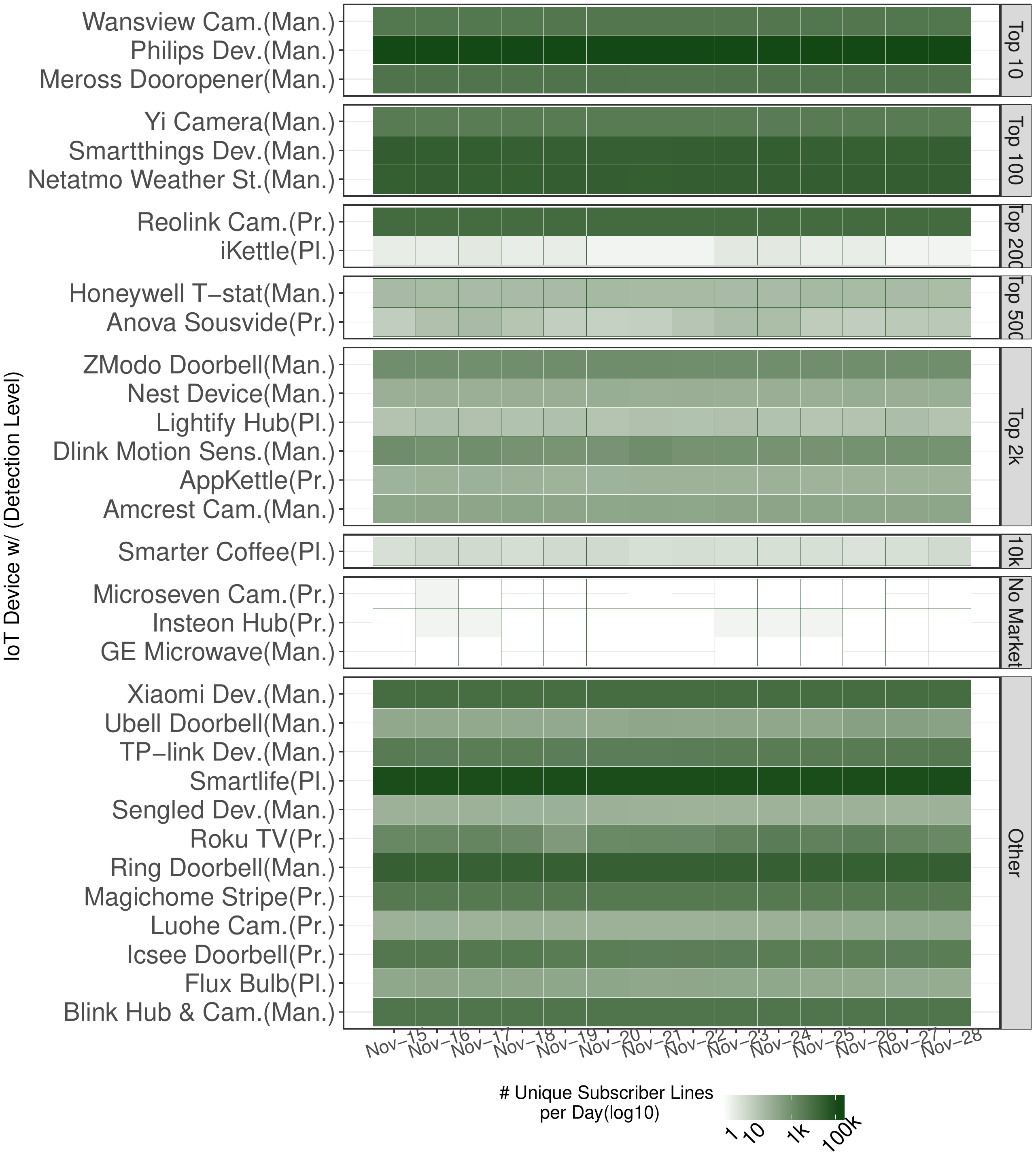}
	\caption{ISP: Drill down of IoT activity for 32 different IoT device types with their popularity in the ISPs country.}
	\label{fig:iot-33devs-wild}
\end{figure}

\begin{figure*} [t]
	\centering
	\captionsetup{skip=.5em, width=0.48\linewidth,font=small}
	\begin{minipage}[t]{0.47\linewidth}     
		\includegraphics[width=1\linewidth]{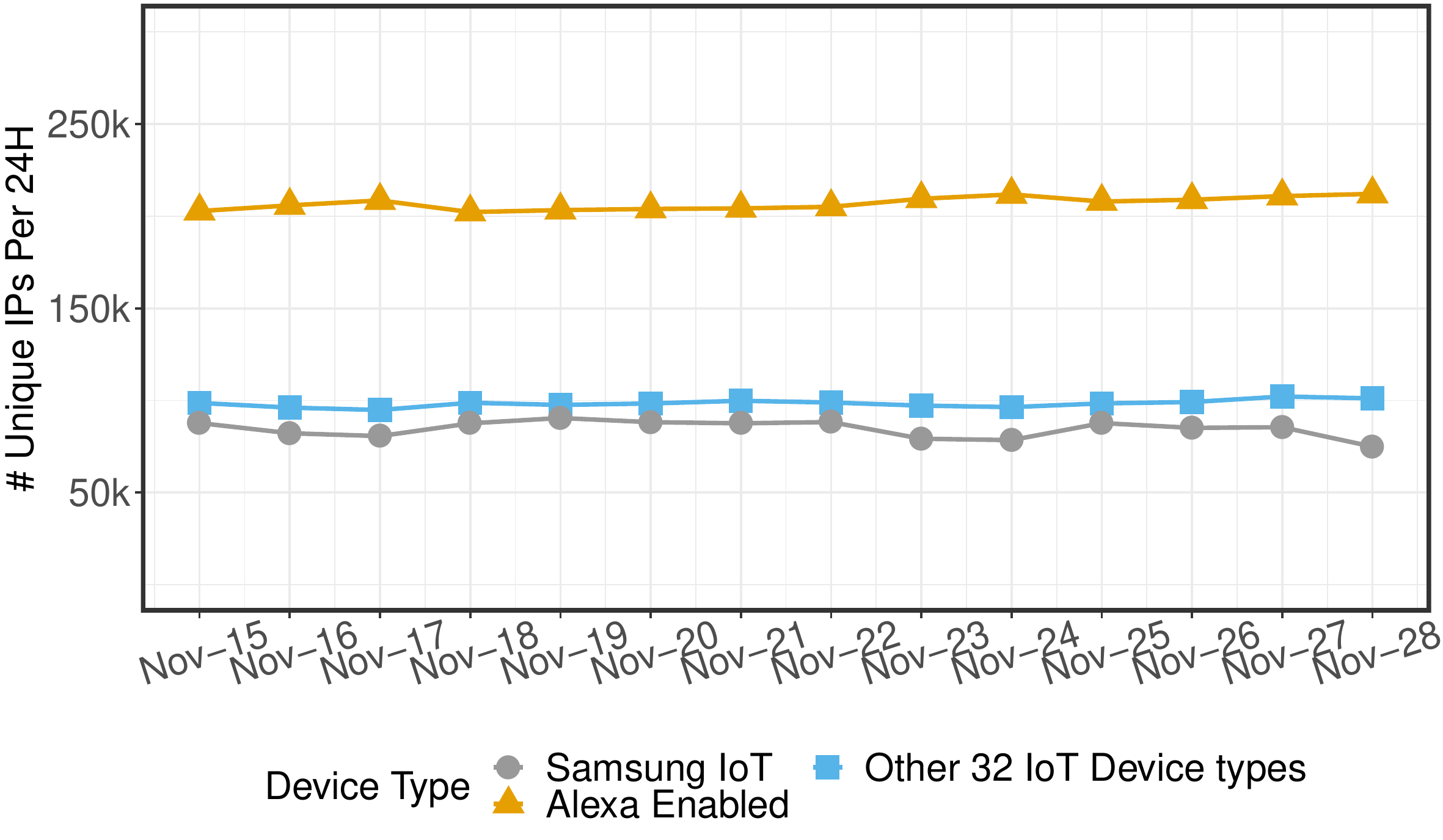}
		\caption{IXP: Number of Samsung IoT, Alexa Enabled, and Other 32 IoT device types IPs observed/day.}
		\label{fig:all-devs-daily-ixp}
	\end{minipage}
	\hfill
	\hspace{0.2cm}
	\begin{minipage}[t]{0.48\linewidth}     
		\captionsetup{skip=.5em, width=0.95\linewidth,font=small}
		\centering
		\includegraphics[width=1\linewidth]{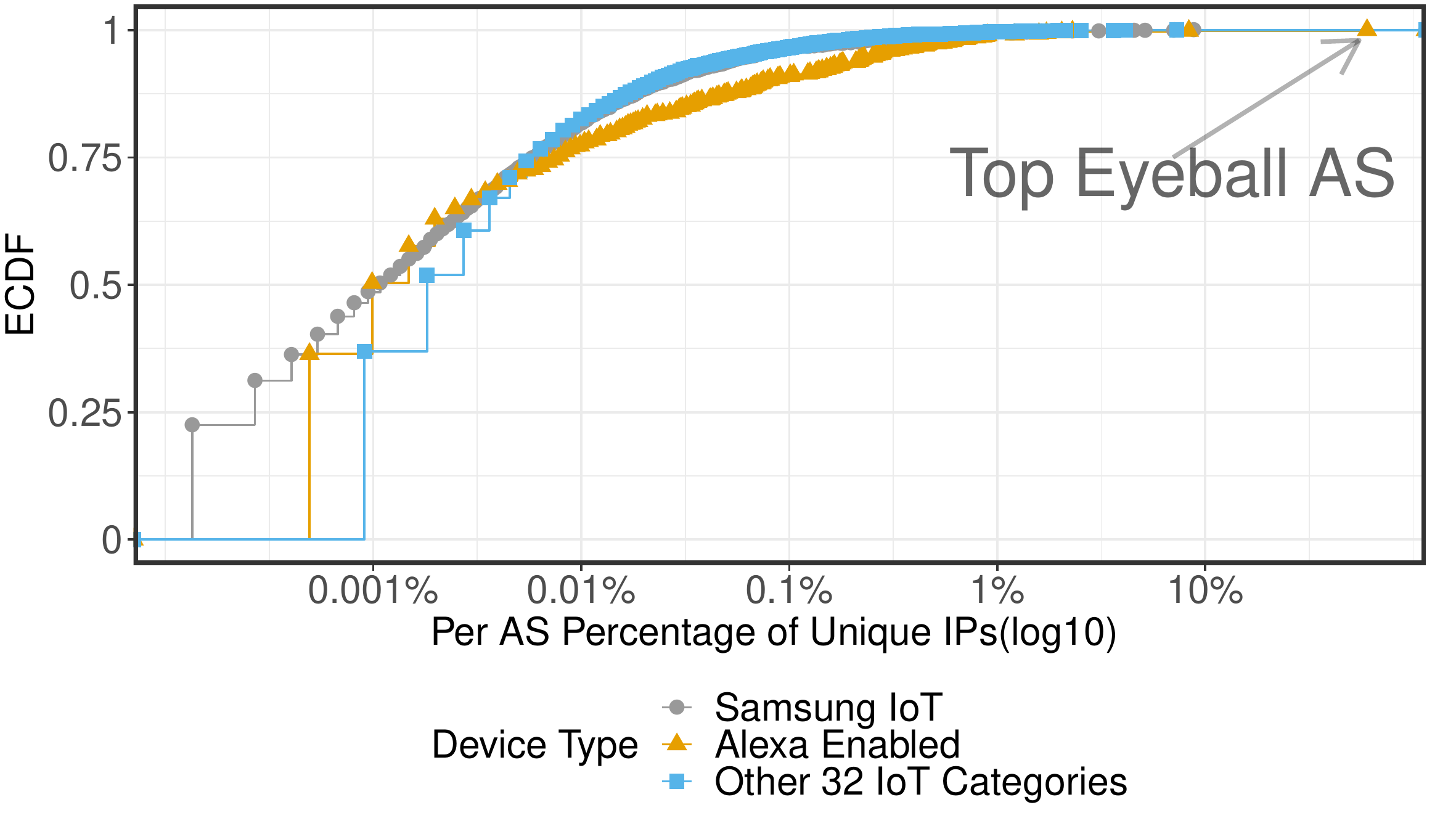}
		\caption{IXP: ECDF of Per-ASN Percentage (\# Unique IPs) - Day 15-11-2020.}
		\label{fig:perasn-percentage-ixp}
		
	\end{minipage}	
\end{figure*}

\para{IoT related activity in-the-wild.} Figure~\ref{fig:alexa-wild-dev-hour-day} shows the number of ISP subscriber
lines for which we detect IoT related activity. \new{The ISP does not operate a carrier-grade NAT. Even if multiple IoT devices are hosted at an ISP subscriber, we count the hosting subscriber only once. Thus, the number of subscribers that host a given IoT device is a lower bound for the number of the given IoT device in the premises of ISP subscribers.} Figure~\ref{fig:alexa-samsung-wild-dev-hour} and Figure~\ref{fig:alexa-samsung-34dev-wild-dev-day} focus on hourly and daily summaries.  Since the top IoT devices
detected are Alexa Enabled and Samsung IoT, we show them separately. 
We see IoT related activity for roughly 20\% of the subscriber lines. 
Our results show a significant penetration of Alexa Enabled devices of roughly 14\%. This is slightly more than
estimates of national surveys in the country where the ISP operates, stating that the market penetration of Alexa
Enabled devices, as of June 2019, is around 12\% ~\cite{bitkomspeakers,alexamarketshare, dwspeakers}. Yet, these reports
cannot capture which devices are in active use at any particular day, e.g., Nov. 2019, contrary to our study. \new{Note,
in Figures~\ref{fig:alexa-wild-dev-hour-day},~\ref{fig:alexa-sam-detailed-day},~\ref{fig:iot-33devs-wild} and~\ref{fig:all-devs-daily-ixp} we apply our
methodology on each time bin independently.} 

\noindent\textbf{Daily patterns of IoT related activity.} By looking at the hourly plots in Figure~\ref{fig:alexa-samsung-wild-dev-hour}, we see some significant daily patterns
for Alexa Enabled and Samsung IoT devices.  We do not see diurnal patterns for the other 32 IoT device types.
Such diurnal patterns are correlated with human activities.  Typically, during the day, network activity increases as
the users interact with the IoT devices while it decreases during the night when the devices are idle.  As detection
likelihood is correlated with network activity, the devices detectability also correlates with this diurnal pattern.  We
note that the patterns for Alexa Enabled does not differ from those for Samsung.  The reason is that many of the Alexa Enabled and Samsung IoT (Samsung TVs) class may be used more for entertainment, which is why their activity is higher in the evenings. Samsung IoT devices have a small spike in the mornings before gradually reaching their peak around 18:00 (ISP timezone).

For the drill down for Samsung IoT devices see Figure~\ref{fig:alexa-sam-detailed-day}.
Even with the presence of a diurnal variation for Alexa Enabled, there is a significant baseline during the night.  
This is expected as IoT devices often have traffic even when they are idle and are thus detectable.  
Over the course of a day, the diurnal variation is rather low compared with the typical network activity driven by human activity. 
This explains the low variance of the observed number of subscriber lines for Alexa Enabled devices.

\noindent\textbf{Aggregation per day.} We observed in Section~\ref{sec:crosscheck} that, while it is often possible to detect Alexa Enabled devices within an hour, the same is not always true for Samsung IoT devices. 
Therefore, Figure~\ref{fig:alexa-samsung-34dev-wild-dev-day} reports the same data but this time using an aggregation period of a day.\footnote{Most subscriber lines are not subject to new address  assignments within a day. Most addresses remain stable as the ISP offers VoIP services.} 
We see that the total number of observed subscriber lines does not change drastically from day to day. 
However, we also note that the number of subscriber lines with Alexa Enabled devices roughly doubled, while those with Samsung increased by a factor of 6. 
The reason is that detecting Samsung IoT devices is more challenging because they are contacting their Primary domain less frequently than Alexa Enabled devices. Thus, their detection is heavily helped by the increase in the observation time period. 
For the other IoT devices we see these effects, whereby the increase is correlated to the expected time for detection.
Note, certain Samsung domains are contacted by both Samsung IoT and Non-IoT devices. In our analysis, we only consider domains that are exclusively contacted by \emph{Samsung IoT} devices. By adding those domains, the number of detected Samsung devices will be increased at least by a factor of two, but this also adds false positives to our results.

\noindent\textbf{Detecting specific devices.} So far, we have focused on the superclass of Alexa Enabled and Samsung IoT devices. 
However, by adding more specialized features, our methodology allows us to further differentiate them. 
For example, some subsets of domains are only contacted by specific products. 
Thus, in Figure~\ref{fig:alexa-sam-detailed-day} we show which fraction of the Alexa Enabled IoT devices are confirmed Amazon products and which fraction of these are Fire TVs using a conservative detection threshold of 0.4. 
For Samsung IoT devices, we show how many of them are Samsung TVs. 
Again, the number of subscriber lines with such IoT devices is quite constant across days. 
As expected, the specialized devices only account for a fraction of the devices of both manufacturers. 

\noindent\textbf{Subscriber lines churn.} While the ISP's overall churn of subscriber line identifier is pretty low 
\new{(as was also confirmed by the ISP operator)}, some changes are possible and may bias our results.  
Possible reasons for such changes are: unplugging/rebooting of the home router, regional outages, or daily re-assignment of IPs for privacy reasons. 
Yet, as most IoT devices are detectable within a day (recall Section~\ref{sec:crosscheck}), the churn should not bias our results. 
Still, to check for such artifacts, we move to larger time windows: see the upper panel of Figure~\ref{fig:alexa-wild-cumulative-day}, which plots the cumulative number of subscriber lines with detected Alexa Enabled and Samsung IoT devices, respectively, for up to two weeks. 
Here, we see that the fractions increase. 
However, we may have substantial double counting due to identifier rotation. 
To underline this conclusion, we consider penetration at the /24 prefix aggregation level, see the lower panel in Figure~\ref{fig:alexa-wild-cumulative-day}. 
The penetration lines stabilize smoothly, but at different levels and with different speed. 
The latter is related to the popularity of an IoT device. 
If it is already popular, the likelihood of moving from a known to an unknown subscriber line identifier is lower with respect to less popular IoT devices.
\begin{figure*} [t]
	
	\begin{minipage}[t]{0.5\linewidth}
		\captionsetup{skip=.5em,font=small}
		\includegraphics[width=1\linewidth]{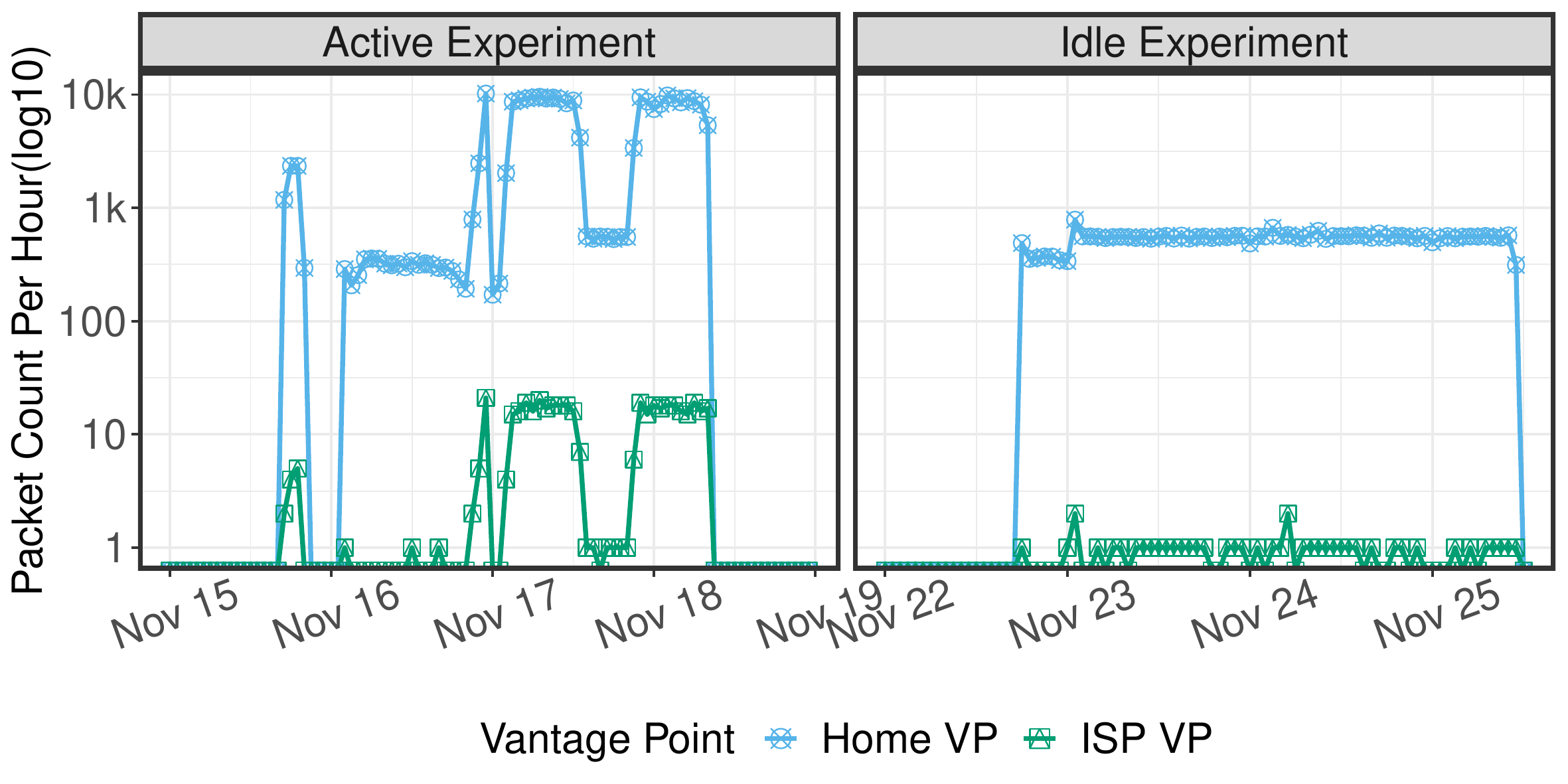}
		\caption{Home-VP/GT Household: Single Alexa Enabled device.}
		\label{fig:single-alexa-enabled-packet-count}
	\end{minipage}
	\hfill
	\begin{minipage}[t]{0.46\linewidth}
		\captionsetup{skip=.5em, font=small}
		\includegraphics[width=1\linewidth]{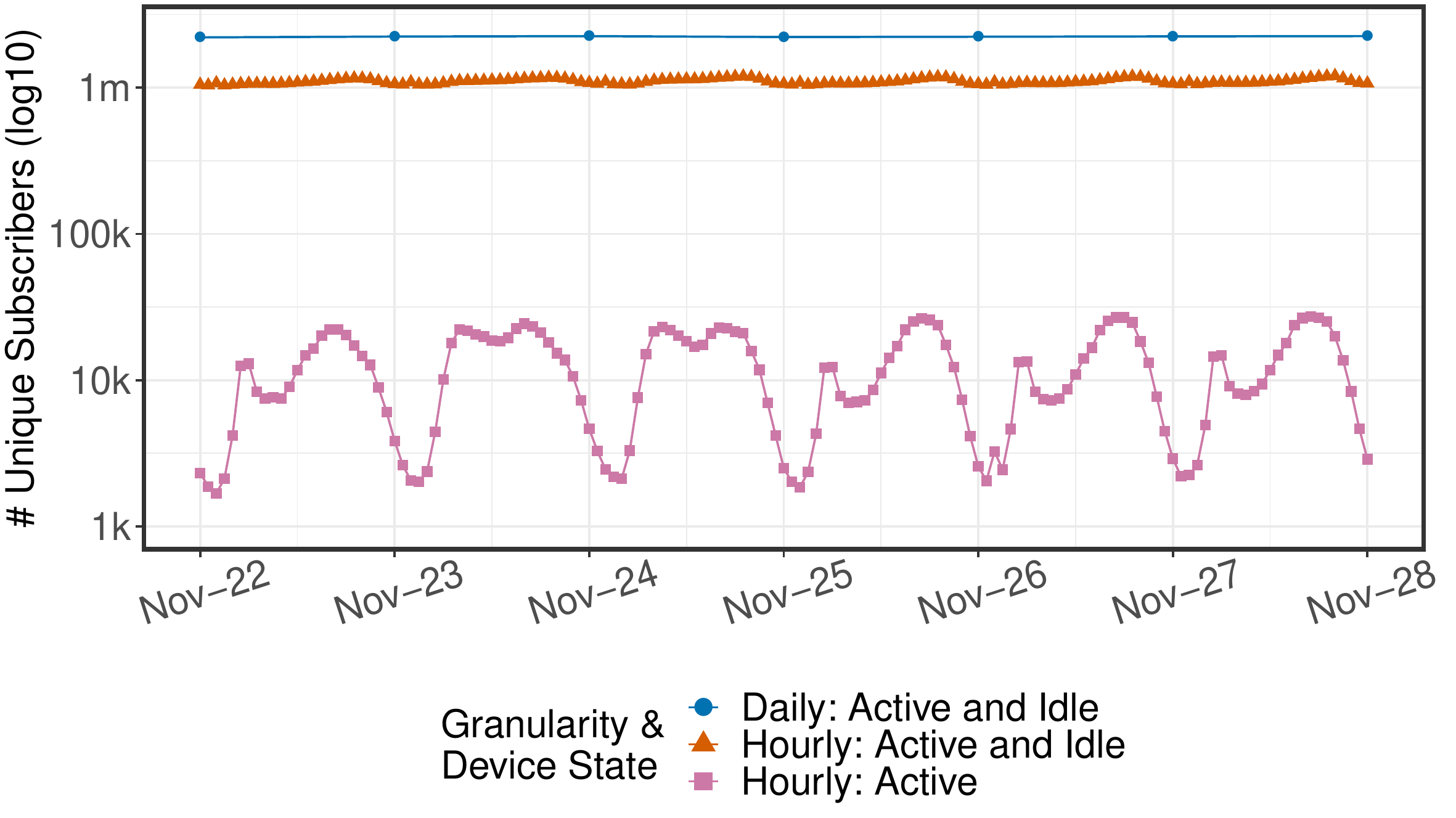}
		\vspace{-0.3cm}
		\caption{ISP: \# Subscribers with active Alexa Enabled/hour.}
		\label{fig:alexa-enabled-packet-count}
	\end{minipage}
\end{figure*}

\noindent\textbf{Detecting other IoT devices in-the-wild.} Figure~\ref{fig:iot-33devs-wild} reports the detected number of the IoT devices that are neither Alexa Enabled nor Samsung IoT. We report them using a heatmap,
where each column corresponds to a day and each row to an IoT device annotated with its detection level.  
The color of each entry shows the number of subscribers lines during that day.
Our first observation is that the number of subscriber lines for each device class is very stable across the duration of our study.
Next, we point out that our experiments include popular devices from both the European as well as the US market. 
For a reference, we report the relative popularity of each IoT device in the Amazon ranking for that device, in the country where the ISP operates. If a ranking of a device is not available, we categorize them as ``other.''   
Popular devices are more prominent than unpopular ones or the ones that are not available in the country's market.  
For example, on the one hand there are Philips devices that are popular and in heavy use with more than 100\,K subscription lines on a daily basis. 
On the other hand there is Microseven camera that is not in the country's market. 
Yet, we can still observe some deployments, these results highlight that our methodology is able to detect both popular and unpopular IoT devices when the domains and associated service IPs that IoT devices visit can be extracted. 

\subsection{Vantage Point: IXP}\label{sec:IXP-results}

Next, we apply our detection methodology at the IXP vantage point. Here, we have to tackle a few additional challenges: First, the sampling rate at the IXP is an order of magnitude lower than at the ISP. 
Second, the vantage point is in the middle of the network, which means that we have to deal with routing asymmetry and partial visibility of the routes. 
Third, while the ISP does aggressive spoofing prevention, e.g., with reverse path filtering, this is not possible at the IXP.  
Spoofing prevention is the responsibility of individual IXP members. 
Thus, we require TCP traffic to see at least one packet without flags, indicating that a TCP connection was successfully established. 
While this may reduce visibility, it prevents us from over-estimating the presence of IoT traffic.

While the IXP offers network connectivity for every ASes, only a few member ASes are large eyeballs~\cite{rasti2010eyeball}.
It is not that surprising that we did not observe any activity of the ground truth experiment, recall Section~\ref{sec:visibility}.
Still, we are able to detect significant IoT activity.
 Figure~\ref{fig:all-devs-daily-ixp} shows the number of IPs for which we detected IoT activity per day for our two-week study period (November 15th-28th, 2019). 
 We are able to detect roughly 90k Samsung devices, 200k Alexa Enabled devices, and more than 100k of other IoT devices. 
 This underlines that our methodology, which is based on domains and generalized observations from a single subscriber line, is successful.
\eat{We observe a noticeable decrease in the number of IPs for Samsung IoT devices between November 16th-23rd, 2019. 
Manual investigation showed that the traffic of some member ports responsible for a significant fraction of the Samsung IoT devices dropped significantly during this time period. 
The reason for this decrease is that the local government of these IXP members restricted access to the Internet in the wake of political demonstrations~\cite{internetblockirannetblock, internetblockiranwashpost}. 
After the end of this local shutdown the traffic activity was back to its previous level and with that also our number of detected Samsung IoT devices. This highlights the sensitivity of our methodology.
}
Most IXP members are non-eyeball networks. 
As such, we expect that the detected IoT activity is concentrated on these members. 
Figure~\ref{fig:perasn-percentage-ixp} shows an ECDF of the distribution of IoT activity per AS for one day (November 15th, 2019) and three IoT device types, namely, Samsung IoT,  Alexa Enabled, and the other IoT devices. 
The distributions are all skewed---a small number of member ASes are responsible for a large fraction of the IoT activity. 
Manual checks showed that these are all eyeball ASes. 
Yet, we also see a fairly long tail. 
This underlines that some IoT devices may not only be used at home (and, thus, send their traffic via a non-eyeball AS).


\section{Discussion}\label{sec:discussion}

\subsection{Device Usage Detection}

A natural question is whether sampled flow data also allows one to distinguish if an IoT device is in active use. 
Our results indicate that the answer is positive. 
First, our ground truth experiments show  that for some devices, the domain sets used during the idle experiments differ from those during active experiments. 
Hence we can use these domains to determine the mode (active/idle) of an IoT device. 
Second, the amount of traffic also varies depending on the mode. 
To highlight this, Figure~\ref{fig:single-alexa-enabled-packet-count} shows the number of observed packets at the Home-VP for a single Alexa Enabled device, as well as the ISP-VP for both modes. 
Activities cause spikes above 1K at the home vantage points and above $10$ at the ISP-VP. 
These ranges are never reached during the idle experiments.

When using the first insight for, e.g., devices from TP-link (TP-link Dev.), we are able to capture active use for only 3.5\% of the devices. 
The reason is that these are plugs, which have a total traffic volume so low that it limits the detectability due to the low sampling rate at the ISP.
When using the second insight for Alexa Enabled devices, we find that we can detect significant activity.  
Figure~\ref{fig:alexa-enabled-packet-count} shows both the subscriber lines with Alexa-enabled devices per hour, per day as well as the subscriber lines with active Alexa-enabled devices. Based on the above-mentioned observations, we used the threshold of $10$ for packet counts per hour to filter out subscribers that actively used Alexa-enabled devices in a given hour. Based on this threshold, we see that the number of actively used devices reaches 27,000 during the day and weekends (November 23rd-24th, 2019), following the diurnal pattern of human activity.   

The ability to distinguish active from idle usage of IoT devices in the wild may raise ethical/privacy concerns. 
However, the goal of this paper is not to analyze user behavior, but rather to point out the privacy concerns associated with having these IoT devices at home~\cite{PETS20_DuboisKMPCH20}.

\subsection{Potential Security Benefits}

The ability to detect IoT services can be used in a constructive manner or even as a service by ISPs. 
For example, if there are known security problems with an IoT device, the ISP/IXP can block access to certain domains/IP ranges or redirect their traffic to benign servers. 
The methodology can also be used for troubleshooting, incident investigation, and even incident resolution.
For example, an ISP can use our methodology for redirecting the IoT devices traffic to a new backend infrastructure that offers privacy notices or security patches for devices that are no longer supported by their manufacturers.

Moreover, if an IoT device is misbehaving, \eg if it is involved in network attacks or part of a botnet~\cite{USENIXSS17_AntonakakisABBBCDHIKKLMMMSSTZ17}, our methodology can help the ISP/IXP in identifying what devices are common among the subscriber lines with suspicious traffic.
Once identified, their owner can be notified in a similar manner, as suggested by ~\cite{NDSS19_CetinGLAKITTYv19}, and  it may be possible to block the attack or the botnet control traffic~\cite{SPWTCP29_MandalariKHDC20}. 

\subsection{Limitations}

Our methodology has some limitations. 

\noindent\textbf{Sample devices.} We need to have sample devices in order to observe which domains are being contacted. 

\noindent\textbf{Superclass detection.} We mostly check for false negatives and  limitedly for false positives as we only have traffic samples from a subset of IoT devices, but not for all possible IoT devices.
If an IoT device relies on a shared backend infrastructure or common IoT APIs, we only detect the superclass, \eg at the manufacturer level. 

\noindent\textbf{Network activity.} We rely on the network activity of IoT devices. 
As such, if the traffic volume is very low detectability decreases, and detection time increases. 

\noindent\textbf{Shared infrastructures.} We cannot detect IoT services that rely on shared infrastructures. If the IoT devices change their backend infrastructure, \eg after an update, we may have to update our detection rules too.

\subsection{Lessons Learned}

Our analysis could be simplified if an ISP/IXP had access to all DNS queries and responses \new {as they do in~\cite{guo20} and~\cite{iotfinder}}.  
Even having a partial list, \eg from the local DNS resolver of the ISP, could improve our methodology. 
Yet, this raises many privacy challenges. 
An increasing number of end-users rely on technologies like DNS over TLS~\cite{Google-DNS-over-TLS}, or public DNS resolvers, \eg Google DNS,
OpenDNS, or Cloudflare DNS, rather than the local ISP DNS server~\cite{Akamai-Mapping-EDNS-2015}.
Yet, this also points to another potential privacy issue---the global data collection and analysis engines at these DNS operators, which can identify IoT devices at scale from the recorded DNS logs using our insights. 
Capturing DNS data from the network itself would require deep packet inspection and thus, specialized packet capture, which is beyond the scope of this paper.

The subscriber or device detection speed varies depending not only on the device and its traffic intensity, but also on the traffic capture sampling rates. 
The lower this rate, the more time it may take to detect a specific IoT device. 
Moreover, identifying the relevant domains for each IoT device does require sanitization, which may involve manual work, \eg studying manuals, device documentation,
vendor web sites, or even programming APIs. 
Given that we are unable to identify IoT services if they are using shared infrastructures (\eg CDNs), this also points out a good way to hide IoT~services. 

\new{
\subsection{Future Directions}
We can use our insights to develop signatures that allow an ISP to identify
households that use specific IoT services. If such services are, e.g., subject
to security concerns they can use such signatures to notify the corresponding
customer of the potential problem and fix.  This is also possible if the IoT
service is no longer supported or needs end-user manual upgrades, e.g., to
mitigate threats.  Such signatures may also be used to move from DDoS attacks
towards identifying culprits. 
Our approach is potentially scalable further using MUD profiles~\cite{IETF-RFC8520}, 
where devices will signal to the network what sort of domains, access and network functionality they require to properly function.
It is also possible to extend the list of signatures of IoT devices using crowdsourcing~\cite{sensing-IoT:2019}.}

\section{Related work}\label{sec:related-work}

There have been some recent papers in understanding home IoT traffic patterns and identifying devices based on their signatures, trackers, and network traffic~\cite{DATW16_AptrhorpeRF16}. These approaches often rely on testbed data~\cite{Information-Exposure-IMC2019,TMC19_SivanathanGLRWVS19}, or tools for the active discovery of the household devices and their network traffic~\cite{IoTInspector}. The authors in~\cite{TMC19_SivanathanGLRWVS19} use a broad range of network features from packet captures, including domain names to train a machine learning model and detect IoT devices in a lab environment. However, they do not further study the backend infrastructure supporting IoT devices. There have also been a few early attempts at mitigating against these device discoveries using traffic padding~\cite{PPET19_ApthorpeHRNF19} or blocking techniques~\cite{SPWTCP29_MandalariKHDC20}. 

A number of recent efforts focused on inferring IoT device types from network traffic~\cite{LCN19_SivanathanGS19, AUDI-IoT-JSAC}. In~\cite{CIOTDI20_MazharS20} the authors used instrumented home gateways to look at IoT traces from over 200 households in a US city. Their analysis revealed that while the IoT space is fragmented, few popular cloud and DNS services act as a central hub for the majority of the devices and their data. 

\new{Generally, many IoT devices periodically connect to specific servers on the Internet. Authors in~\cite{iotfinder} and~\cite{guo20} proposed a method to identify IoT devices by observing passive DNS traffic and unique IP addresses that the device connects to. Unfortunately, many IoT devices rely on shared infrastructures and often different IoT devices from the same vendor connect to the same servers, therefore detection at the scale of ISP/IXP, based on the IP addresses and port numbers without considering the important role of shared infrastructures, cannot be very reliable.}

Complementing the approaches based on testbeds and home gateways, there have been efforts in understanding IoT traffic patterns using data from transit networks~\cite{IoT-TransitNets2020}, though it has been challenging to successfully validate the derived signatures. Similar works relied on specific port numbers~\cite{port_scan} that may also be used for specialized industrial IoT systems~\cite{IoT-Industrial2020}, though the approach used cannot be easily extended to general-purpose IoT devices and  smart home systems that utilize popular ports, e.g., 443, 80.

These related works indicate that often, neither data from core networks subject to sampling and middleboxes, nor data from few devices using home gateways or testbeds are enough for rapidly and accurately detecting IoT devices, and understanding their anomalies and misconfigurations~\cite{haddadi2018siotome}. 

In this paper, for the first time we have complemented detailed ground truth data from testbeds and a particular subscriber, with large-scale data from an ISP and an IXP, to reveal the aggregate behavior of these devices, alongside the ability to isolate and identify specific subscriber devices using sampled data at an ISP.

\section{Conclusion}\label{sec:conclusion}

Home IoT devices are already popular, and their usage is expected to grow
further. Thus, we need to track their deployment without deep packet inspection
or active measurements, both intrusive and unscalable methods for large
deployments.  Our insight is that many IoT devices contact a small number of
domains, and, thus, it is possible to detect such devices at scale from sampled
network flow measurements in very large networks, even when they are in idle mode. We
show that our method is able to detect millions of such devices in a large
ISP and in an IXP that connects hundreds of networks.

Our technique is able to detect 4 IoT platforms, 20 manufacturers and 11 products--both
popular and less popular ones--at vendor level and in many cases even at
product granularity. While this detection may be useful to understand the
penetration of IoT devices at home, it raises concerns about the general
detectability of such devices and the corresponding human activity.  

In light of our alarming observations, as part of our future work, we would like to
investigate how to minimize the harm of potential attacks and surveillance
using IoT devices. 
We also want to use our insights to help ISPs to tackle security and
performance problems caused by IoT devices, e.g., by detecting them,
redirecting their traffic, or blocking their traffic.

\section*{Acknowledgements}

\new{We thank the anonymous reviewers and our shepherd Kensuke Fukuda for their constructive feedback. This work was supported in part by the European Research Council (ERC) Starting Grant ResolutioNet (ERC-StG-679158), the EPSRC Defence Against Dark Artefacts (EP/R03351X/1), the EPSRC Databox (EP/N028260/1), and the NSF (CNS-1909020).}
\vspace{0.2cm}

\balance

%
%
%
%

\bibliographystyle{unsrt}
\interlinepenalty=10000
\bibliography{paper}



\end{document}